\documentclass[a4paper,fleqn]{cas-dc}

\usepackage[numbers]{natbib}
\usepackage{epsfig}
\usepackage{float}
\usepackage{graphicx}
\usepackage{soul}

    \usepackage{framed} 
    \usepackage{multicol} 
     
    \usepackage{nomencl} 
    \makenomenclature
    \renewcommand*\nompreamble{\begin{multicols}{2}}
    \renewcommand*\nompostamble{\end{multicols}}

\begin{document}
\let\WriteBookmarks\relax
\def\floatpagepagefraction{1}
\def\textpagefraction{.001}
\shorttitle{Optimisation of crust freezing in meat processing via Computational Fluid Dynamics}
\shortauthors{E. Greiciunas et~al.}

\title [mode = title]{Optimisation of crust freezing in meat processing via Computational Fluid Dynamics}                      



\author[1]{Evaldas Greiciunas}


\address[1]{School of Mathematical Sciences, University of Nottingham, NG7 2RD, UK}
\author[1]{Federico Municchi}
\author[2]{Nicodemo Di~Pasquale}
\address[2]{School of Mathematics and Actuarial Science, University of Leicester, LE1 7RH, UK}
\author[1]{Matteo Icardi}
\cormark[1]




\cortext[cor1]{Corresponding author}


\begin{abstract}
In this work  a numerical model for two-dimensional axisymmetric continuous freezing by impingement of processed meat or similar products in food industry moving along a conveyor belt is presented.  The model represents  a more computationally efficient alternative to solve conjugate heat transfer between a fluid and a solid, accompanied by phase change in some constituents of the solid phase.
In the model presented here  it is assumed that the solid can be represented as an homogeneous medium, with  its thermophysical properties depending on the temperature. The  impingement freezing model is conceived to be valid for highly processed vegetarians products or meat such as sausages, mince or ham freezing. 
Furthermore, this approach is much simpler in terms of computational cost whilst it still captures the  complexity of continuous freezing under industrial setting.
The methodology is implemented as a new solver in the widely used open-source Computational Fluid Dynamics (CFD) library OpenFOAM\textsuperscript{\textregistered}. 
Overall, highly non-linear freezing behaviour was found due to the phase change inside the solid and the associated heat of fusion. 
We studied the effect of high fluid Reynolds numbers as well as investigating the optimal distance between the jet and the solid surface for different speeds of the conveyor.
We found that the maximum freezing is obtained positioning the jet at a distance H = 7.2D (where D is the diameter of the impinging jet) and setting the speed of the conveyor such that the P\'eclet number of the solid is $\text{Pe}_{\text{s}}=8244$. 
The methodology developed allows to obtain detailed insight on the freezing process for various impingement configurations at a minimum computational cost using a freely available open-source tool.
\end{abstract}


\begin{highlights}
\item Computational model for continuous impingement freezing was developed
\item Continuous solid (burger meat) domain undergoes phase-change
\item Parametric study shows non-linear freezing behaviour  
\item The results show a potential of highly optimising the impingement freezing performance  
\end{highlights}


\maketitle

\begin{table*}
\begin{framed}
\printnomenclature
\end{framed}
\end{table*}
\nomenclature{h}{Heat transfer coefficient [W/(m$^2$ K)]}
\nomenclature{$\mathbf{u}$}{ Fluid velocity field [m/s]}
\nomenclature{$\mathbf{v}$}{ Conveyor velocity [m/s]}
\nomenclature{$p_{rgh}$}{Pressure head [Pa]}
\nomenclature{$h_{\text{f}}$}{Fluid enthalpy [J/Kg]}
\nomenclature{$h_{\text{s}}$}{Solid enthalpy [J/Kg]}
\nomenclature{$T$}{Temperature [K]}
\nomenclature{$\mu$}{Fluid viscosity [Pa s]}
\nomenclature{$t_{fz}$}{Dimensionless freezing time}
\nomenclature{$x_{fz}$}{Axial frozen coordinate [m]}
\nomenclature{$y_{fz}$}{Radial frozen coordinate [m]}
\nomenclature{$\tau_{\text{eff}}$}{Effective shear stress [Pa]}
\nomenclature{$\mu_{\text{eff}}$}{Effective viscosity [Pa s]}
\nomenclature{$y^{+}$}{Shear wall distance}
\nomenclature{$\dot{q}_{fs}$}{Interface heat flux [W/m$^2$]}
\nomenclature{$T_{in,\text{f}}$}{Jet temperature at the inlet [K]}
\nomenclature{$T_{sf}$}{Mean temperature at the fluid-solid interface [K]}
\nomenclature{Re}{Reynolds number of the jet}
\nomenclature{$U_{in}$}{Jet inlet velocity [m/s]}
\nomenclature{$\text{Pe}_\text{s}$}{Peclet number of the solid}
\nomenclature{$\alpha_{\text{eff}}$}{Effective thermal diffusivity of the fluid  [m$^2$/s]}
\nomenclature{$\alpha_{\text{s}}$}{Thermal diffusivity of the solid  [m$^2$/s]}
\nomenclature{$Cp$}{Specific heat  [W/(m$^2$  K)]}
\nomenclature{$\kappa$}{Thermal conductivity  [W/(m K)]}
\nomenclature{$\rho_{\text{f}}$}{Fluid density  [kg/m$^3$]}
\nomenclature{$\rho_{\text{s}}$}{Solid density  [kg/m$^3$]}
\nomenclature{$Fr$}{Frozen crust}
\nomenclature{H1}{Solid radius [m]}
\nomenclature{D}{Jet diameter [m]}
\nomenclature{L}{Domain length [m]}
\nomenclature{H}{Jet-solid distance [m]}
\nomenclature{$\tau_w$}{Wall shear stress [Pa]}
\nomenclature{IF}{Impingement Freezing}
\nomenclature{CFD}{Computational Fluid Dynamics}
\nomenclature{HPF}{High Pressure-assisted Freezing}
\nomenclature{HF}{Hydrofluidisation Freezing}
\section{Introduction}

\noindent An important part of food industry is represented by freezing of produce such as vegetables and meat. Food freezing is a complex problem which needs to take into account several different parameters for a complete description of the process, such as freezing time, food quality, and freezing cost. 
Historically, commercial freezing of food products was obtained through cryogenic immersion or mechanical freezing \cite{Agnelli2001,James2015}. 
A combination of the two above mentioned processes can also be used: firstly, through cryogenic immersion a  rapid formation of a protective layer of the food is accomplished, which functions both
as a protection  during transportation and prevents losses in the moisture content when subject to slow  mechanical freezing \cite{Agnelli2001}.
Cryogenic freezing typically uses liquid $\text{N}_2$ or liquid $\text{CO}_2$ \cite{Soto} and has the highest rate of heat transfer compared to other processes due both to the high temperature gradients between the coolant and the food and because of the evaporation of the refrigerant (latent heat of vaporisation). 
Additionally, the high rates of heat transfer reached in cryogenic freezing result in the formation (nucleation) of smaller ice crystal inside the solid. This is associated with higher food quality since the ice crystals produced by cryogenic freezing are too small to damage the food structure  \cite{spiess1980impact,Marazani2017, Kaale2011,poulsen1977freezing}. 
However, cryogenic freezing has two main disadvantages: firstly, the sudden freezing induces stresses in the products which can lead to damage \cite{Zhou2010}. Secondly it is an economically expensive technique because of the volumes of cooling liquid required  (up to $1\,  \text{kg}$ of $\text{N}_2$ per $1 \, \text{kg}$ of processed product) \cite{Salvadori2002,Soto}). 
Mechanical freezing is cheaper than cryogenic freezing, however, it is less  efficient (heat transfer coefficients  $h<<50\,\text{W}/(\text{m}^2 \text{K})$). The reduced heat transfer rate  leads to the growth of significantly larger ice crystals and thus to a reduced of quality of the final product \cite{Salvadori2002}.  
As a result, there is a substantial interest for developing alternative fast low cost food freezing techniques such as Impingement Freezing (IF), High Pressure-assisted Freezing (HPF), Hydrofluidisation Freezing (HF), description of which can be found in \cite{James2015, Kaale2011,Marazani2017}. 

\begin{figure*}
    \centering
    \begin{picture}(490,110)
    \put(250,0){\includegraphics[width=0.5\textwidth]{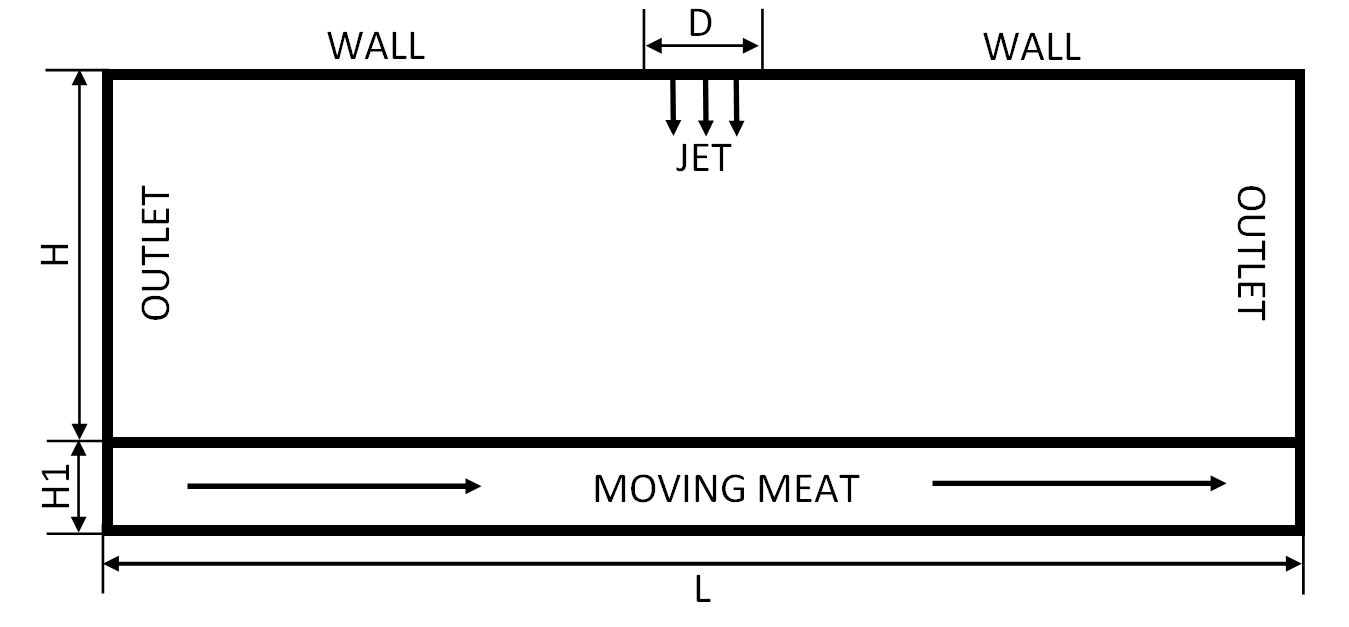}}
    \put(0,0){\includegraphics[width=0.5\textwidth]{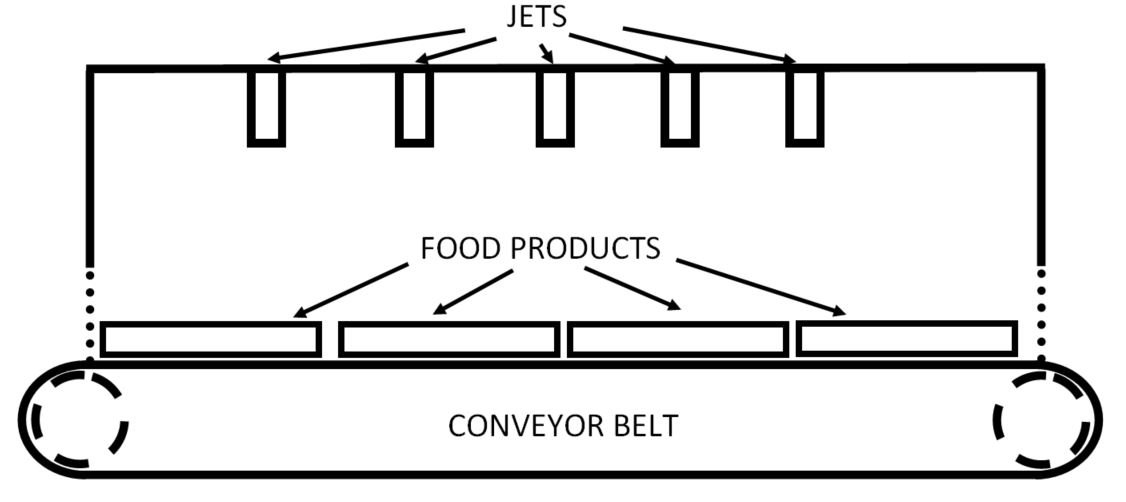}}
    \put(10,100){a)}
    \put(245,100){b)}
    \end{picture}
    \caption{a) Typical impingement freezing setup used in industrial refrigeration \cite{lee1998impingement}. b) Simplified axisymmetric setup for continuous impingement freezing.}
    \label{fig:model}
\end{figure*}

Impingement freezing is essentially an improved mechanical freezer \cite{James2015} in which cold air jet is perpendicularly directed towards the food.
This results in an enhanced heat transfer due to the break-up of the fluid boundary layer next to the solid surface \cite{newman2001cryogenic}. 
Traditionally, an impingement freezer for food industry would have the following components (Figure \ref{fig:model}a) \cite{Salvadori2002}:
\begin{itemize}
    \item Freezing chamber.
    \item Grid or conveyor belt on which the produce is placed and transported. 
    \item One or multiple nozzles which supply high speed cooling air. Nozzles can be installed perpendicularly to the belt or at different angles. Additionally, some nozzles can be placed along the conveyor belt to supply air at different temperatures \cite{lee1998impingement,Kaale2011} 
\end{itemize}
It is worth to notice that impingement freezing is one of few new techniques which have been fully commercialised \cite{sarkar2004modeling,Sarkar2004,winney2012field} due both to its cost effectiveness compared to cryogenic freezing, and to the significantly lower freezing times compared to conventional mechanical freezing. 
Clearly, air based impingement works best for dense food products with high surface area (since air does provide an efficient heat transfer). However, the process is well suited for rapid surface freezing applications (such as crust freezing) due to its capabilities for fast freezing  \cite{James2015}.
It was also shown that for some applications, impingement freezing is able to produce similar freezing time compared to cryogenic freezing (e.g. for small burgers\cite{sundsten2001effect}) without the complexity of the cryogenic process.   
Furthermore, IF is  62-79\% faster and with 36-72\% reduced weight loss when compared to conventional freezing, thanks to the highest heat transfer coefficients\cite{Salvadori2002}. 
An experimental study by \cite{Soto} showed that the heat transfer coefficient ranges between $70-250$ $\text{W}/(\text{m}^2\text{K})$, depending on the regime of the cooling air.
However, it should also be noted that whilst increasing jet velocity reduces the freezing time it could also have damaging effects to the structure of the food \cite{Soto}.  
Additional drawbacks of the IF with respect to standard freezing equipment are: the higher  installation costs and power consumption. However, these are offset by much faster product processing capabilities \cite{winney2012field}. 

IF is more complex than conventional mechanical freezing both from  food product and fluid perspectives, and there is a significant interest in optimising this procedure. 
The optimisation of this process can be obtained by either experiments or numerical modelling, since analytical relations can only be found for excessively simplified cases.
Generally, in experimental works the heat transfer coefficient is measured for both control samples \cite{anderson2006effective} or real food \cite{Soto} products under impingement conditions. 
However, such measurements are difficult and experiments tend to be expensive.  
An interesting scenario was considered by \cite{sarkar2004modeling}, who investigated the optimum jet placement. They found that the best freezing conditions can be obtained by placing the jets at approximately 6-8 jet diameters away from the freezing surface. 

Numerical modelling studies  usually focus their attention mainly on the food procude domain \cite{Salvadori2002,Agnelli2001,Pham2006,erdogdu2005mathematical,pham2014freezing}. 
In these studies, the main objective is the description of the solid freezing process, its various associated parameters such as the mass diffusion, important for porous products (e.g. bread), and its related processes such as recrystallisation \cite{Pham2006}. 
The single solid domain modelling approach requires special boundary conditions, either coolant temperature or a heat transfer coefficient at the boundary corresponding to a certain freezing process. These boundary conditions  cannot accurately define complex cooling process of an impinging jet. 
The computational complexity from modelling both fluid and solid domains resulted in studies which addressed the problem only  under certain limiting hypothesis.
\cite{olsson2004heat} and \cite{Dirita2007} examined the effect of impinging jet cooling of a cylindrical food product placed on a conveyor belt (Figure \ref{fig:model}a) using Computational Fluid Dynamics (CFD) from a frontal view perspective.   
In both cases turbulent air was modelled using the $k\textnormal{-}\omega$ $\text{SST}$ model, which is effective in capturing near-wall effects \cite{menter2003ten} with the results showing highly non-linear heat transfer coefficient along the solid surface. 
However, in the case of \cite{olsson2004heat} only the solid boundary was modelled whilst \cite{Dirita2007} used a Conjugate Heat Transfer (CHT) formulation with no phase change. 
The phase change was omitted purely based on arguments of numerical stability, since the resulting sudden change in thermophysical properties can lead to difficult convergence \cite{Pham2006}. This, combined with the non-linear nature of fluid dynamics makes the modelling challenging.  
An attempt to model both the impinging jet and the solid cooling in axisymmetric coordinates was reported by \cite{jafari2008analysis}. However, the lack of numerical details, the computing resolution (12000 cells maximum in total) and overall mesh quality raise some questions regarding the quantitative accuracy of this study.

In this work, a continuous axisymmetric impingement freezing model with CHT is developed for food products. 
Contrary to current studies, phase change in the solid is modelled using thermophysical properties of burgers obtained from \cite{Agnelli2001}. 
The model provides a good computational compromise between complexity of fluid dynamics calculations and the phase changing of the solid and is able to predict freezing of continuous dense foods such as sausages, cooked ham, mince.
Additionally, the  model allows tuning of multiple parameters such as jet diameter, jet distance from food, food velocity whilst taking into account complex freezing and impinging jet processes.

\begin{table}
\begin{tabular}{ccc}
\hline \hline
$\rho_{\text{f}}$ & $1.569$               & kg/m$^3$  \\
$C_p$  & $1002.7$              & J/(kg K)   \\
$\mu$  & $1.467\times 10^{-5}$ & Pa s \\
$\text{Pr}$   & $0.728$               & -           \\ \hline \hline
\end{tabular}
\label{tb:air} 
\caption{Constant properties of air at $T=225$ K. Here $C_p$ is the heat capacity and Pr is the (non turbulent) Prandtl number. Notice that the heat conductivity $\kappa$ is calculated employing the definition $\text{Pr} = \rho_{\text{f}} C_p / \kappa$.}
\end{table}
\section{Numerical model}
\begin{figure*}
    \centering
    \begin{picture}(490,200)
    \put(0,0){\includegraphics[width=1\textwidth]{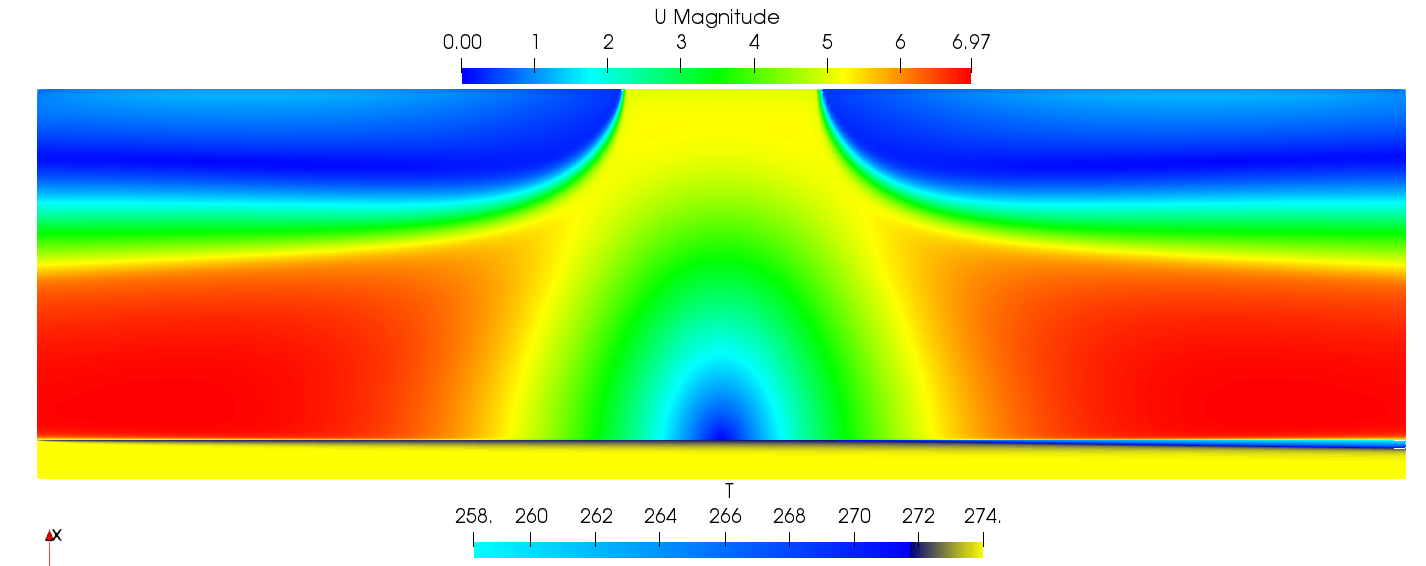}}
    \put(20,115){JET}
    \put(20,32){MEAT}
    \end{picture}
    \caption{Example jet and solid domain visualisation at Fine2 mesh resolution during mesh independence. For the fluid domain - velocity magnitude contours are used whilst for the solid domain temperature contours are used to highlight the freezing front.}
    \label{fig:jet}
\end{figure*}
\subsection{Governing equations for the fluid flow}

\noindent Numerical modelling was performed in OpenFOAM\textsuperscript{\textregistered} using a custom solver with its base built on \verb|chtMultiregionFoam| - a solver for conjugate heat transfer. 
The fluid flow was modelled using RANS (Reynolds Averaged Navier Stokes) equations closed with a $k\textnormal{-}\omega$ $\text{SST}$ turbulence model \cite{chandrasekhar2013hydrodynamic}: 
\begin{align}
\centering
  &\nabla \cdot\left( \rho_{\text{f}} \mathbf{u}\right)=0\\[0pt]
 & \nabla \cdot\left( \rho_{\text{f}} \mathbf{u}\mathbf{u}\right)=-\boldsymbol{\nabla} p_{rgh}+ \boldsymbol{\nabla} \cdot\boldsymbol{\tau}_{\text{eff}} - \left(\mathbf{g}\cdot \mathbf{x}\right)\boldsymbol{\nabla}\rho_{\text{f}} \\[0pt]
&\nabla \cdot\left( \rho_{\text{f}} \mathbf{u}h_{\text{f}}\right)= \boldsymbol{\nabla} \cdot \left( \alpha_{\text{eff}} \boldsymbol{\nabla} h_{\text{f}} \right) - \rho_{\text{f}} \textbf{u}\cdot \textbf{g} + \boldsymbol{\nabla} \cdot \left( \boldsymbol{\tau}_{\text{eff}}\cdot \mathbf{u}\right)
\end{align}
where $\mathbf{u}$ is the fluid velocity field, $p_{rgh}$ is the pressure head, $\rho_{\text{f}}$ is the fluid density, $h_{\text{f}}$ is the fluid enthalpy, $\mathbf{g}$ is the gravitational acceleration, and $\mathbf{x}$ is the spatial coordinate. Furthermore, $\alpha_{\text{eff}} = \alpha_{\text{f}} + \mu_t / (\rho_{\text{f}} \text{Pr}_t)$, where $\alpha_{\text{f}} $ is the effective heat diffusivity of the fluid and $\text{Pr}_t$ is the turbulent Prandtl number taken to be $0.85$ in RANS simulations.
The effective deviatoric stress tensor is given by:
\begin{equation}
    \label{eq::tau}
    \boldsymbol{\tau}_{\text{eff}} = \mu_{\text{eff}}\left( \boldsymbol{\nabla}\mathbf{u} + \boldsymbol{\nabla}^{T}\mathbf{u}\right)
\end{equation}
With $\mu_{\text{eff}} = \mu + \mu_t$, where $\mu$ is the molecular viscosity and $\mu_t$ is the dynamic turbulent viscosity and calculated based on the turbulence model \cite{menter2003ten}.
Wall treatment is undertaken using switchable low and high Reynolds wall functions based on the frictional wall distance:
\begin{equation}
    \label{eq::y+}
    y^{+} = y\frac{\sqrt{\tau_w\rho_{\text{f}}}}{{\mu_{\text{eff}}}}
\end{equation}
Where $\tau_w$ is the wall shear stress and $y$ is the distance between the first cell and the wall. The model switches from laminar to turbulent at $y^{+} =11$ \cite{white2011}. In terms of the fluid properties, incompressible air was used with constant properties at $T=225$ K (summarised in Table \ref{tb:air}). 
In the study, Reynolds number of the jet is defined as:
\begin{equation}
    \text{Re} = \frac{\rho_{\text{f}}U_{in} \text{D}}{\mu}    
\end{equation}

Where $U_{in}$ is the jet inlet velocity and $D$ is the jet diameter (see Figure \ref{fig:model}b). This dimensionless number is employed to represent different working conditions of the impingement device.

\subsection{Governing equations for the solid}

\noindent The solid phase is modelled using an enthalpy based energy conservation equation for moving materials. 
\begin{equation}
\centering
\mathbf{v}\cdot \boldsymbol{\nabla} h_{\text{s}} = \boldsymbol{\nabla} \cdot \left( \alpha_{\text{s}} \boldsymbol{\nabla} h_{\text{s}} \right) 
\end{equation}
Here $h_{\text{s}}$ is the solid enthalpy, $\alpha_{\text{s}}$ is the solid heat diffusivity and $\mathbf{v}$ is the conveyor speed in [m/s],  $\mathbf{v}$ - a constant one-dimensional velocity along the domain axis, resulting in a computationally efficient model representing motion of  solid domain on the conveyor belt.
This additional term is implemented as a programmable source term into OpenFOAM\textsuperscript{\textregistered}. 
Furthermore, the formulation presented here allows to obtain a steady-state solution of a rather complex freezing problem which significantly reduces the computational time. 
Additionally, a custom thermophysical model was implemented using the thermophysical data for burgers reported in  \cite{Agnelli2001}, assuming $70$\% water content throughout the study which results in a freezing temperature of $T_{fz}=271.7$ K. 
This thermophysical model in addition to continuous freezing (allowed by the temperature advection term) enables modelling phase change and the associated heat of fusion effects to the freezing front and heat transfer coefficient.
Finally, a working parameter for the conveyor is defined: the solid P\'eclet number:
\begin{equation}
    \label{eq::Pes}
    \text{Pe}_{\text{s}} = \frac{\text{H1} |\mathbf{v}|}{\alpha_{\text{s}}}   
\end{equation}
Where H1 is the radial length of the solid (see Figure \ref{fig:model}b) and $|*|$ indicates the module operator. This dimensionless number allows to parametrise the working conditions of the conveyor.
\begin{figure}
    \centering
    \begin{picture}(250,305)
    \put(25,0){\includegraphics[width=0.4\textwidth]{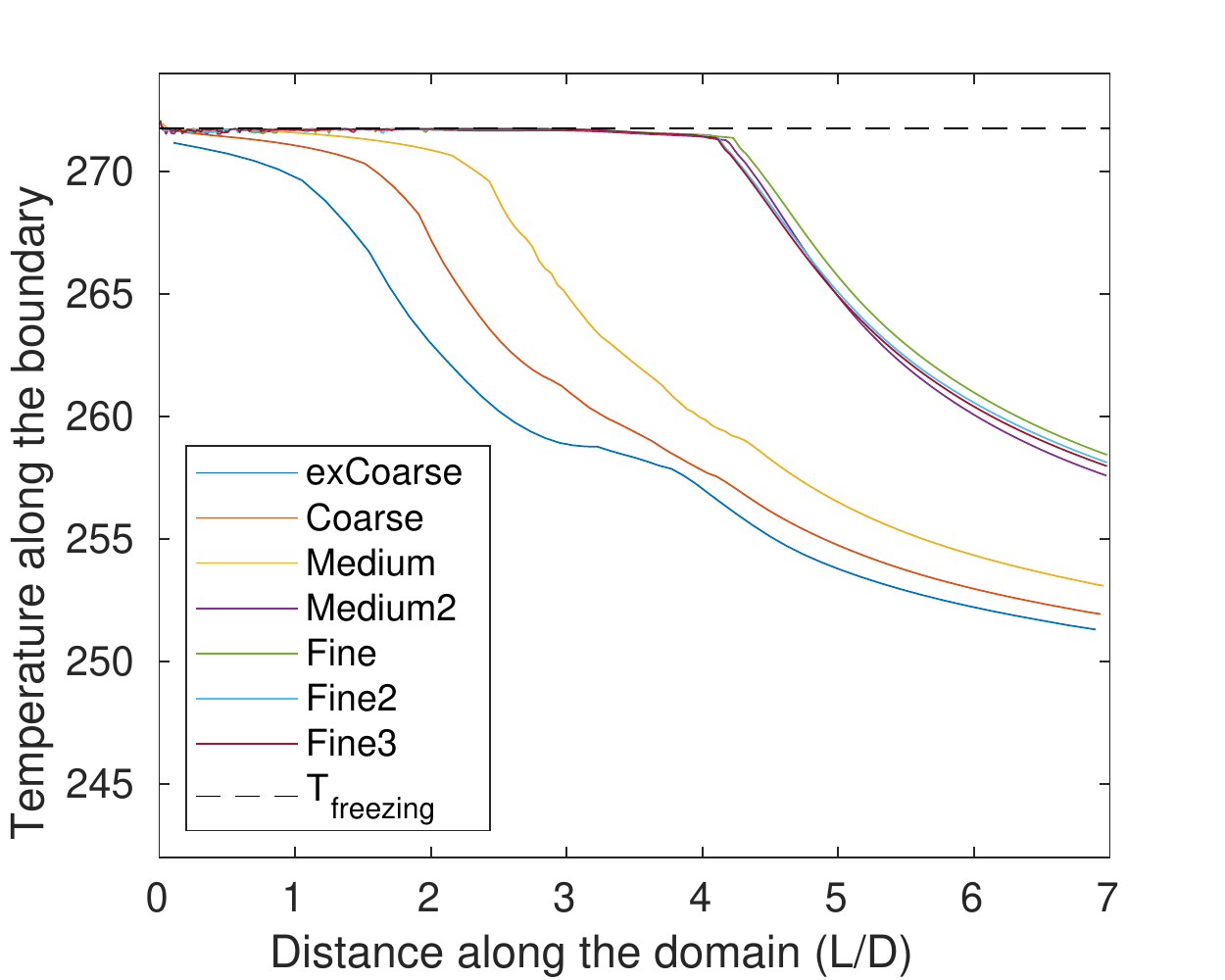}}
    \put(25,155){\includegraphics[width=0.4\textwidth]{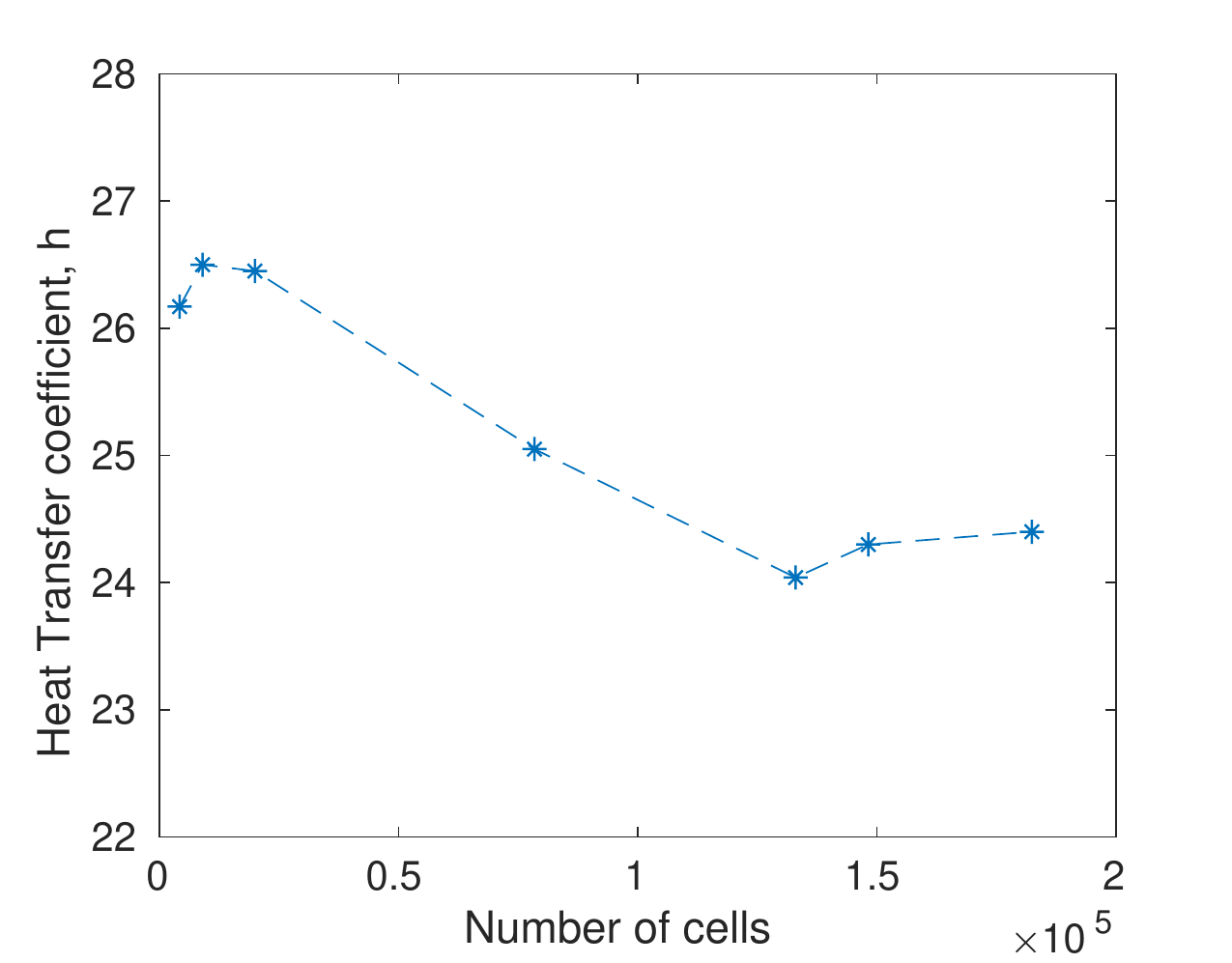}}
    \put(10,280){a)}
    \put(10,140){b)}
    \end{picture}
    \caption{a) Average Heat transfer coefficient h [W/m$^2$ K] at the fluid-solid boundary versus number of cells in different meshes. b) Temperature data along the fluid-solid boundary at various mesh resolutions.}
    \label{fig:independence}
\end{figure}
\subsection{Computational Domain and Grid Independence}

\noindent Computational domain is a simplification of the real process shown in Figure \ref{fig:model}a and is illustrated in Figure \ref{fig:model}b. It is  aimed at studying the non-linear freezing behaviour caused by a jet-solid interaction. The differential operators were discretised in the governing equations employing finite volume formulation and the interpolation schemes shown in Table \ref{tab:schemes}.
Throughout the study, both jet and solid inlet temperatures were kept constant and equal to $T_{in,\text{f}}=225$ K and $T_{in,\text{s}}=274$ K respectively.
The computational domain was built based on the jet diameter D  with dimensions respectively equal to: H1=0.2D, H=1.8D, L=7D.
Numerical grids were built in  OpenFOAM\textsuperscript{\textregistered} using the \emph{blockmesh} utility, which allows the generation of orthogonal hexahedral meshes. 
The main variables of the domain kept were the solid and jet velocities as well as the distance of the jet from the solid, allowing to optimise the freezing process for a variety of scenarios explored throughout this work.
\begin{figure*}
    \centering
    \begin{picture}(490,310)
    \put(25,140){\includegraphics[width=0.4\textwidth]{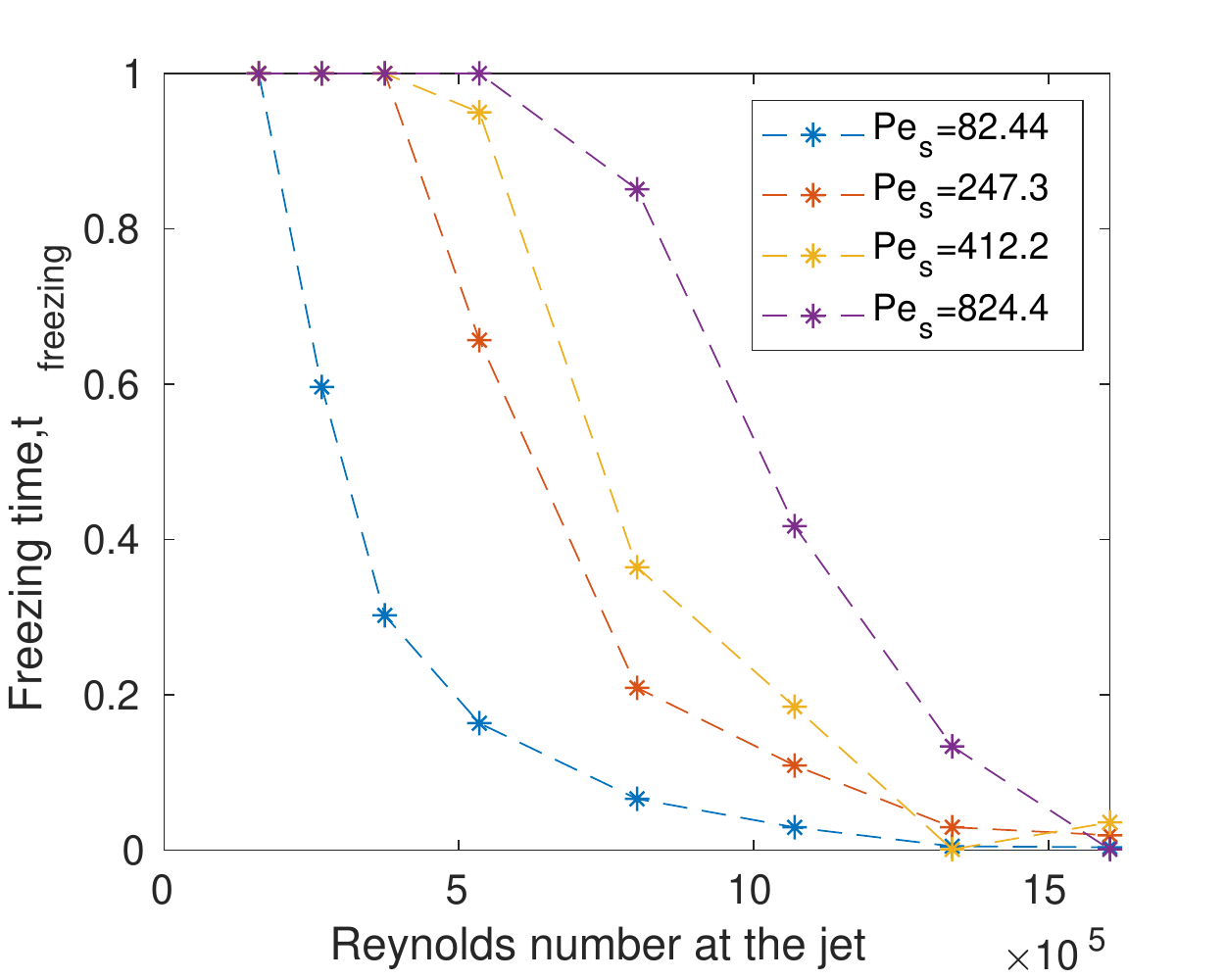}}
    \put(25,0){\includegraphics[width=0.4\textwidth]{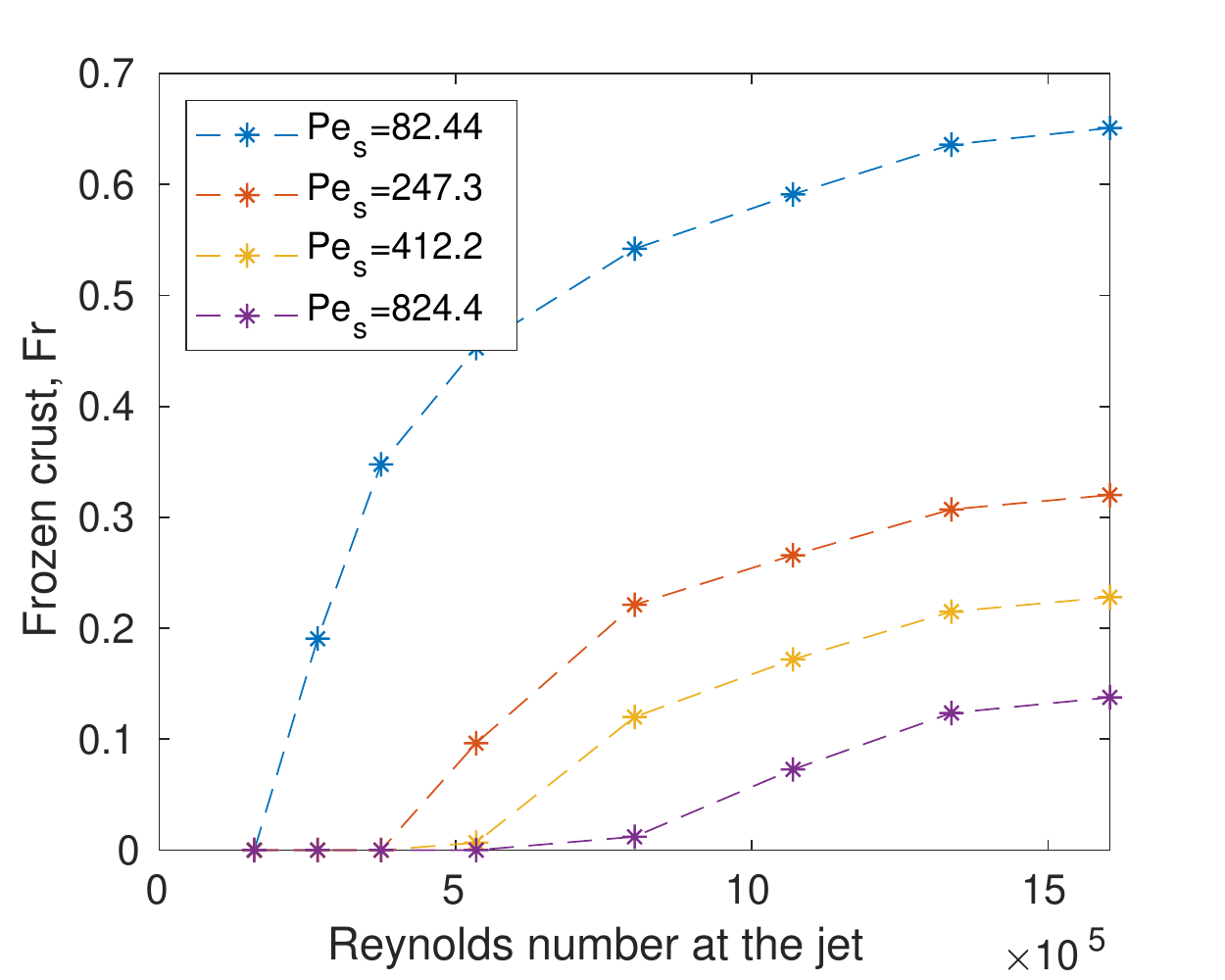}}
    \put(250,140){\includegraphics[width=0.4\textwidth]{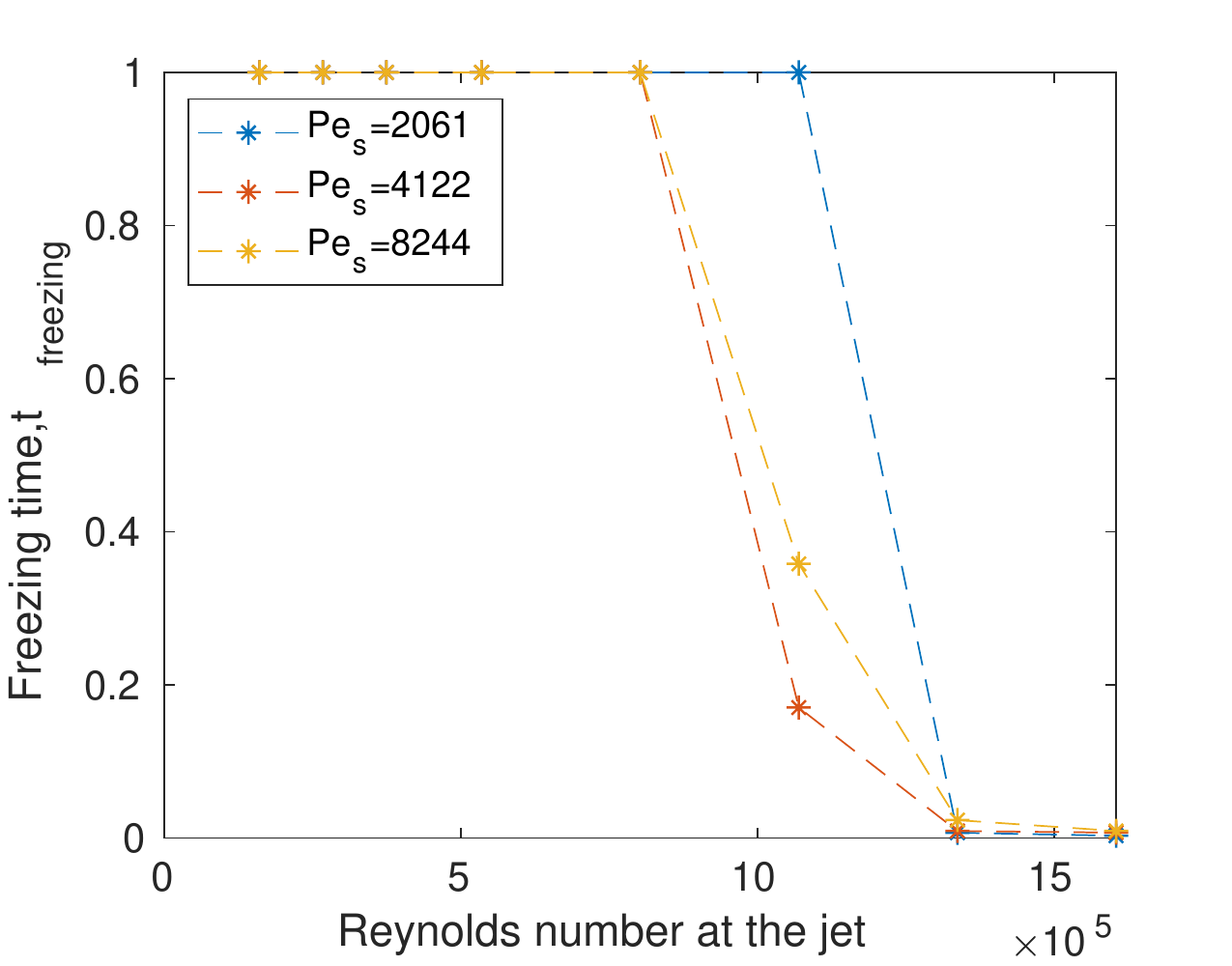}}
    \put(250,0){\includegraphics[width=0.4\textwidth]{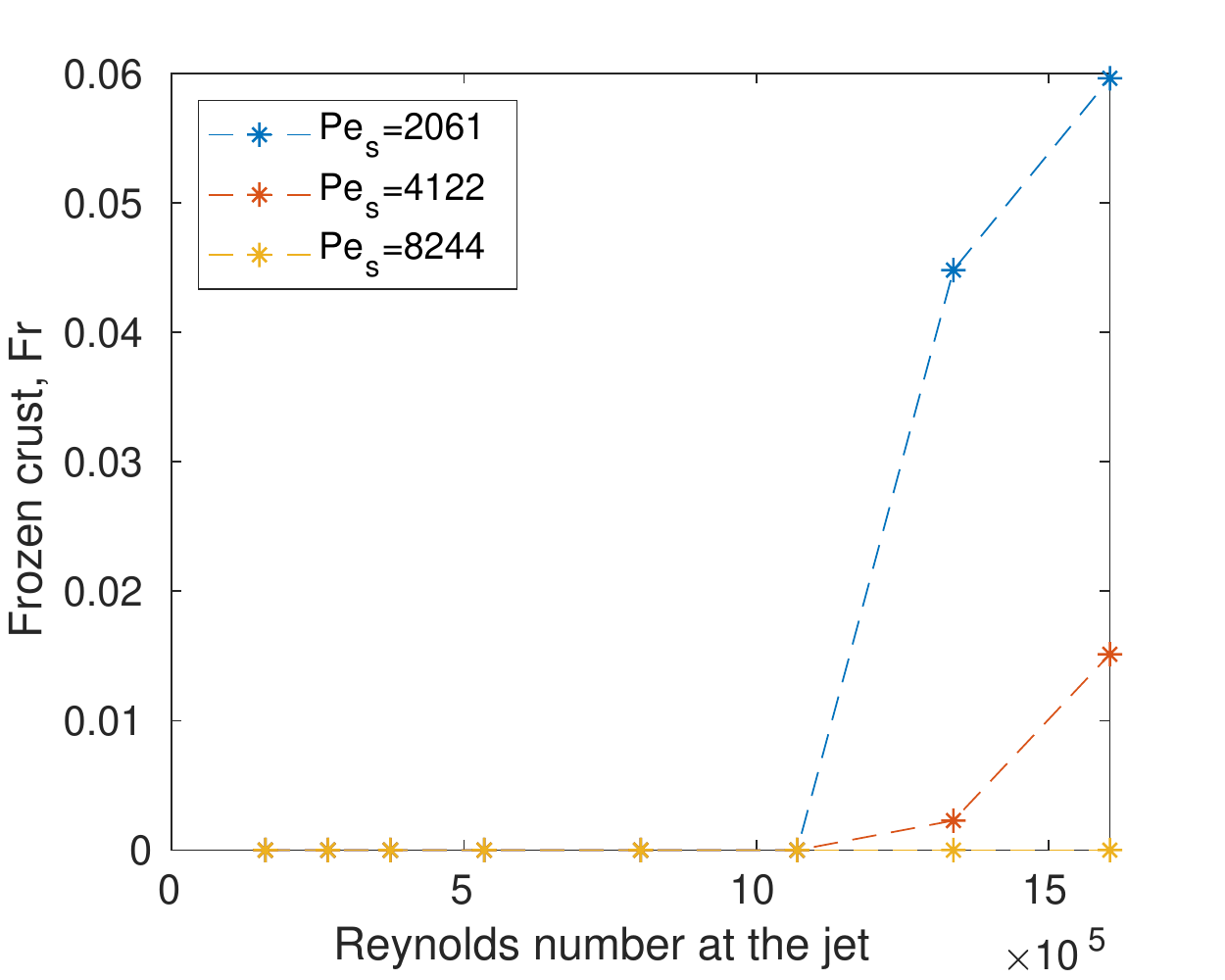}}
    \put(10,280){a)}
    \put(245,280){b)}
    \put(10,140){c)}
    \put(245,140){d)}
    \end{picture}
    \caption{a), b) freezing times associated at various solid Peclet and Reynolds numbers. c), d) Final frozen crust thickness at various solid Peclet and Reynolds numbers. The data is with domain of H=1.8D.}
    \label{fig:freezing_stats}
\end{figure*}
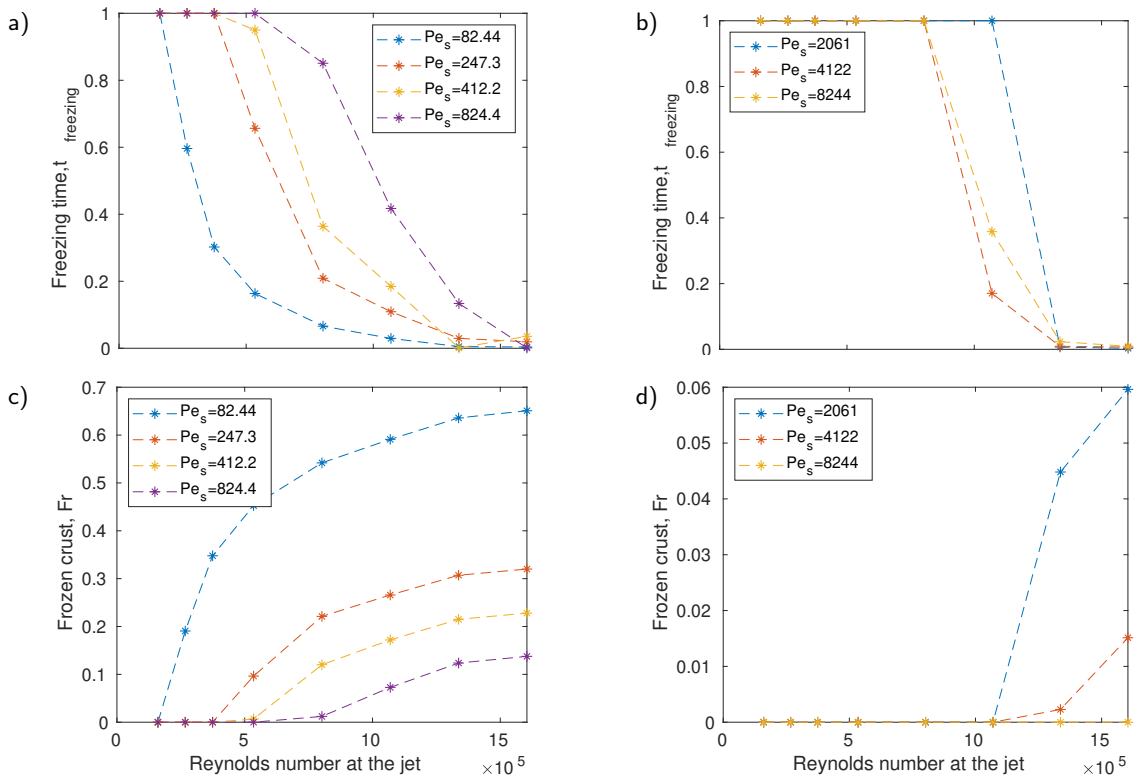
\begin{table}
    \centering
    \begin{tabular}{ll}
    \hline \hline
    gradient ($\nabla$)     &  Gauss linear \\
    laplacian ($\nabla^2$)  &  Gauss linear corrected \\
    div(phi,U)     & Gauss linearUpwindV grad(U)  \\
    div(phi,h) (solid)     & Gauss linearUpwind grad(h) \\
    div(phi,h) (fluid)     & bounded Gauss upwind \\
    div(phi,k)     & Gauss upwind \\
    div(phi,omega) & Gauss upwind  \\ \hline \hline
    \end{tabular}
    \caption{Discretisation schemes as from the \emph{fvSchemes} OpenFOAM\textsuperscript{\textregistered} dictionary used for simulations.}
    \label{tab:schemes}
\end{table}

\begin{table}
    \centering
    \begin{tabular}{lcccc}
            & \multicolumn{3}{c}{Number of cells} & \\ 
         Mesh     & Total  & Solid  & Fluid  & $y+_{mean}$ \\ \hline\hline
         Ex-coarse& 4233   & 663   &  3570   & 21.78 \\
         Coarse   & 9000   & 1500  &  7500   & 15.61\\
         Medium   & 19950  & 3990  &  15960  & 13.38\\
         Medium2  & 78400  & 28000 &  50400  & 0.72\\
         Fine     & 133000 & 49400 &  83600  & 0.08\\
         Fine2    & 148200 & 49400 &  98800  & 0.03\\
         Fine3    & 182400 & 49400 &  133000 & 0.03\\\hline\hline
    \end{tabular}
    \caption{Meshes used for the grid independence study.}
    \label{tab:meshes}
\end{table}

\subsubsection{Grid Independence}

\noindent Sensitivity of the solution was investigated with respect to the mesh resolution using the meshes  reported in Table \ref{tab:meshes}, with operating conditions defined by: $U_{in}=5$ m/s and $|\mathbf{v}| = 1$  mm/s.
These particular operating conditions result in a large amount of solid material becoming frozen, and they therefore are a good indicator of the robustness of the algorithm.
Average heat transfer coefficient $h$ as a function of the grid size was analysed, defined as:
\begin{equation}
    \text{h}=\frac{\dot{q}_{fs}}{T_{in,\text{f}}-T_{fs}}
    \label{eq:h}
\end{equation}
where $\dot{q}_fs$ and $T_{fs}$ represents the total heat exchanged and the average temperature at the interface between fluid and solid respectively. It also should be noted that $h$ is the primary measure for evaluating performance. 
The results reported in Figure \ref{fig:independence}a show a convergence for the values of $h$ for grids finer than the one indicated as \emph{Fine} in Table \ref{tab:meshes}.

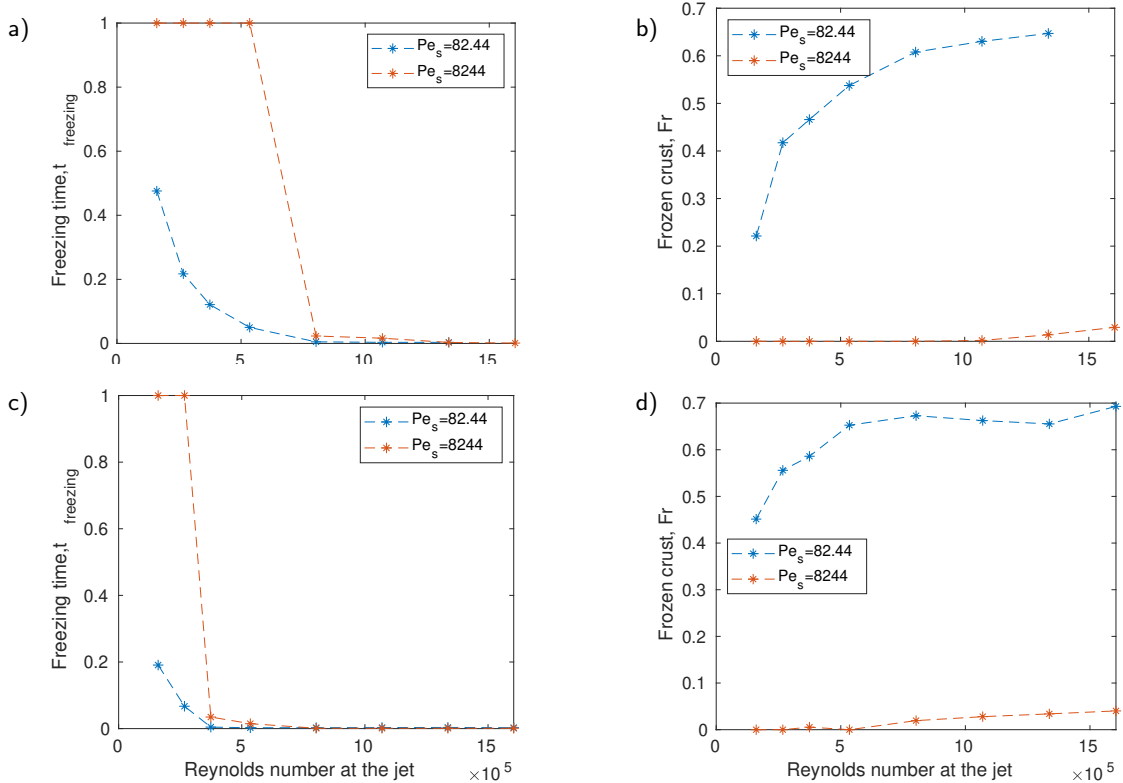
\begin{figure*}
    \centering
    \begin{picture}(490,310)
    \put(25,145){\includegraphics[width=0.39\textwidth]{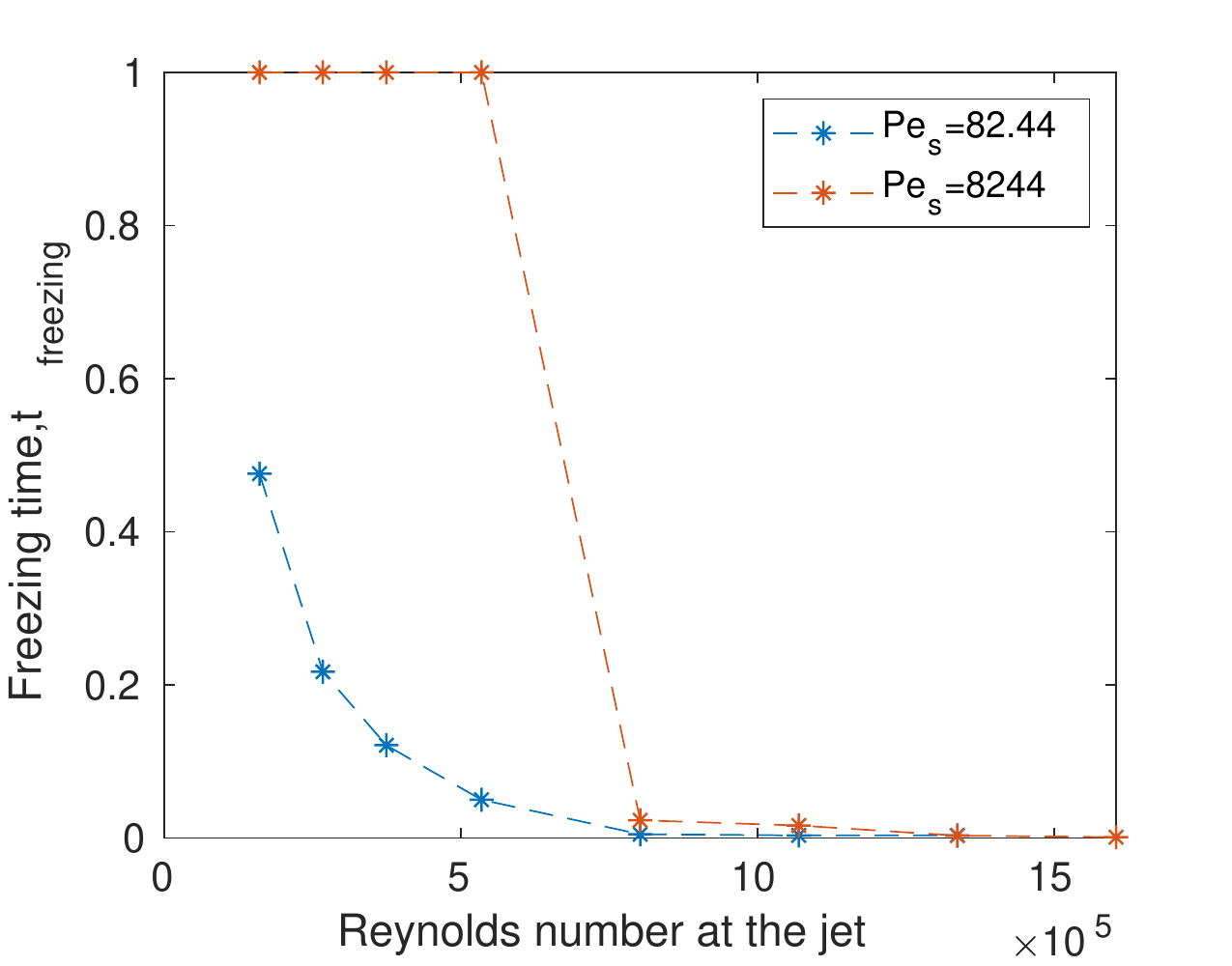}}
    \put(250,145){\includegraphics[width=0.39\textwidth]{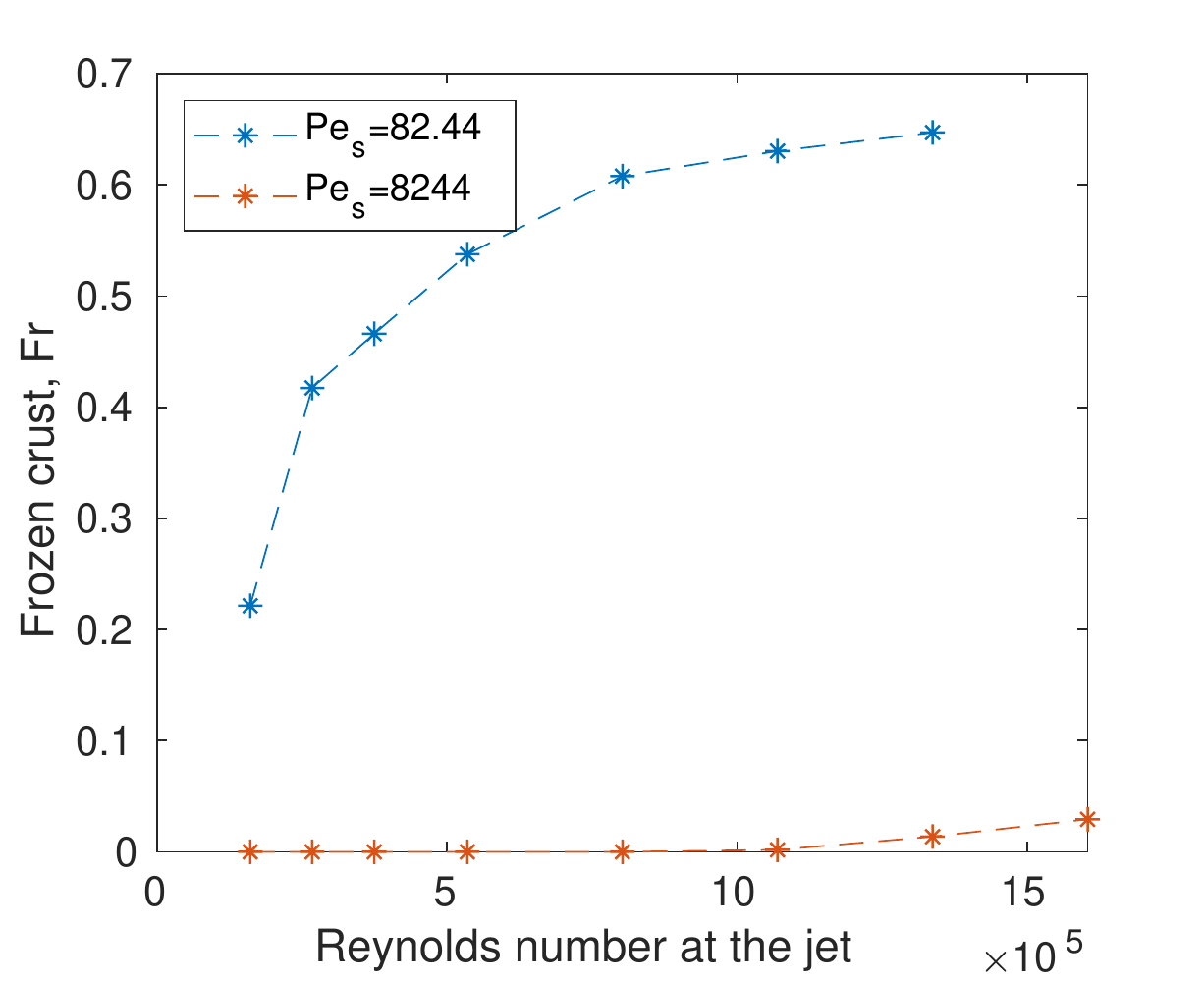}}
    \put(25,0){\includegraphics[width=0.39\textwidth]{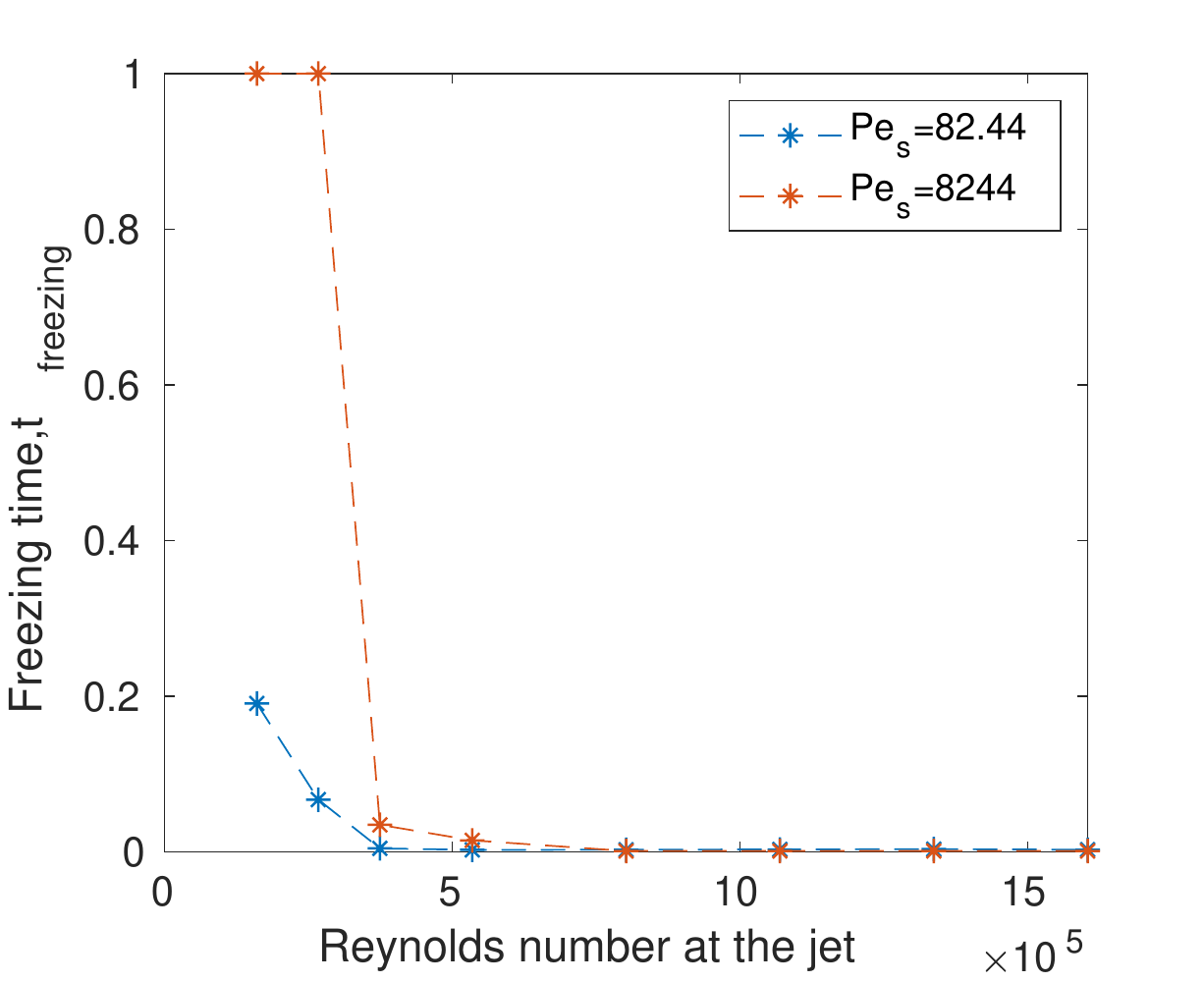}}
    \put(250,0){\includegraphics[width=0.39\textwidth]{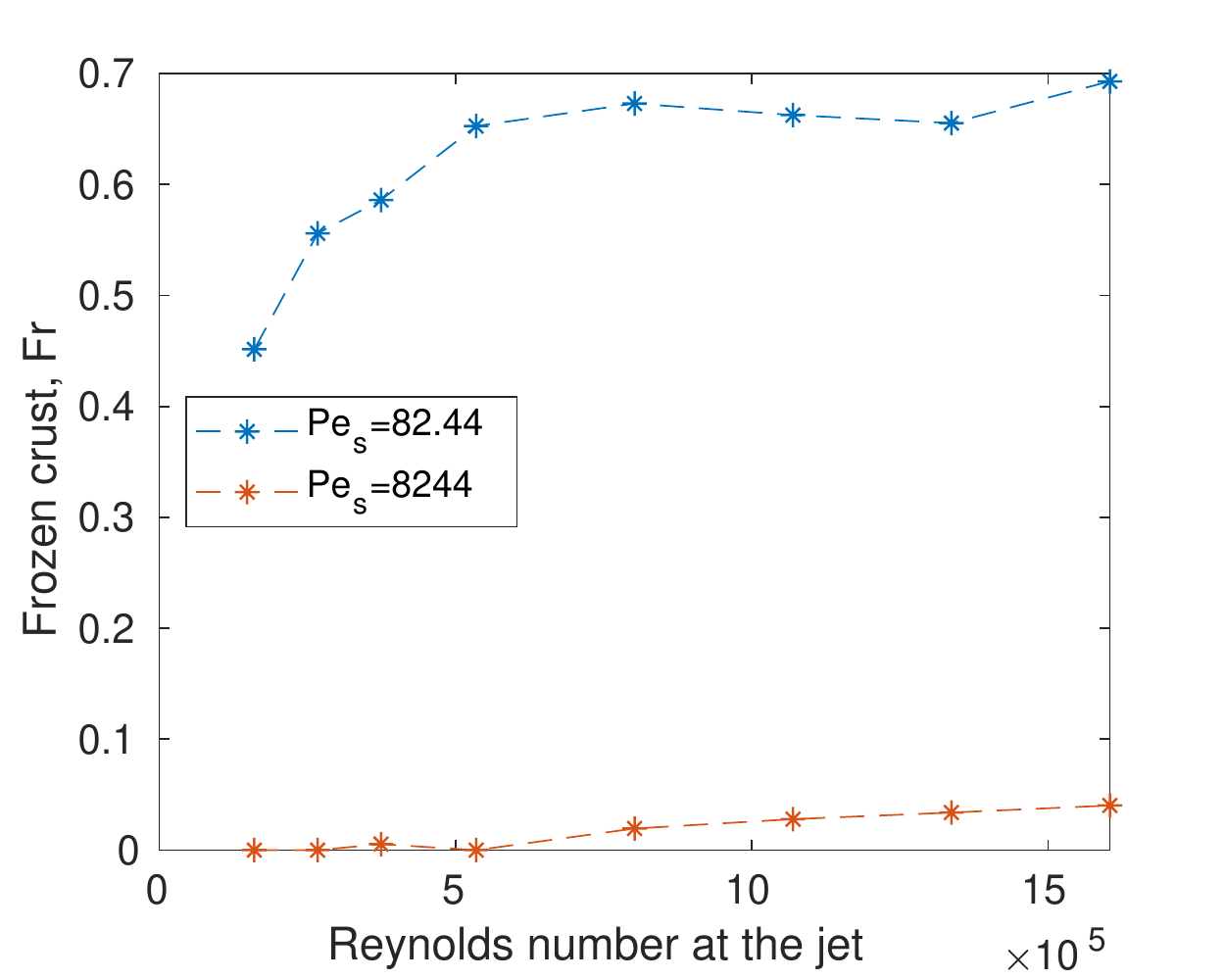}}
    \put(10,280){a)}
    \put(245,280){b)}
    \put(10,140){c)}
    \put(245,140){d)}
    \end{picture}
    \caption{a) Freezing time and b) frozen crust at H=3.6D. c) Freezing time and d) frozen crust at H=7.2D.}
    \label{fig:freezing_stats-jets}
\end{figure*}
Temperature profile along the fluid-solid interface was also considered (Figure \ref{fig:independence}b) which also shows convergence for finer grids.

It can be inferred from Figure \ref{fig:independence}a, that the three finer grids predict a very similar heat transfer coefficient whilst lower resolution grids tend to overestimate the performance of the heat transfer in the model. 
This last observation is also clearly represented in Figure \ref{fig:independence}b, where it is shown that the coarser meshes lead to a rather significant overcooling. 
As a result it was found that the mesh resolution indicated as \emph{Fine2} (see Table \ref{tab:meshes}) was appropriate to conduct the present study. Figure \ref{fig:jet} shows the temperature and velocity fields obtained employing such grid.  
\section{Results}

Parametric study was performed sampling at eight values of Re in the range 
$1.604\times10^5\leq \text{Re}\leq1.604\times10^6$ ( $3\leq U_{in}\leq30$ m/
s) and at seven conveyor speeds in the range $ 1 \leq |\mathbf{v}|\leq 100$ mm/s. 
This choice of parameters resulted in a solid P\'eclet number in the range: $82.44\leq \text{Pe}_\text{s} \leq8244$, allowing to consider a variety of scenarios.

Parameters of interest such as freezing time and the frozen crust thickness were calculated using iso-surface of $T=271$ K in the solid domain ($\approx0.5$ K below a freezing point of the solid). 
The freezing time was calculated by taking the first coordinate of the iso-surface along the axial direction ($x_{fz}$): 
\begin{equation}
\centering
    t_{fz}= \frac{x_{fz}}{|\mathbf{v}| t_{\text{L}}} =\frac{x_{fz}}{|\mathbf{v}|} \frac{|\mathbf{v}|}{\text{L}} =  \frac{x_{fz}}{\text{L}}.
\end{equation}
Note that, in our definition, the freezing time $t_{fz}$ is a dimensionless quantity obtained by dividing it by the time required for the conveyor to perform one passage through the domain $t_{\text{}}=\text{L}/|\mathbf{v}|$.
It should be noted that the initial stable freezing times at low Reynolds numbers in shown Figures \ref{fig:freezing_stats}a  and \ref{fig:freezing_stats}b show that no freezing occurs while the solid is transported through the domain.  

The dimensionless frozen crust thickness ($y_{fz}$) was calculated using the iso-surface radial coordinate at $x=\text{L}$ (Figure \ref{fig:model}b): 
\begin{equation}
    Fr=\frac{y_{fz}-\text{H1}}{\text{H1}}
\end{equation}

Figures \ref{fig:freezing_stats}c and \ref{fig:freezing_stats}d, show a frozen crust of $0$ mm at the lower Reynolds numbers, indicating that the solid did not freeze appreciably, and higher forced convection is required to overcome the latent heat of freezing.
However, these results also show that at the highest Reynolds numbers  the freezing process is almost instantaneous and leads to a significant frozen crust formation in the majority of the situations. 
Interestingly, it can be observed that no significant frozen crust is formed at  $ \text{Pe}_{\text{s}} = 8244$, $\text{Pe}_{\text{s}} = 4122$ and $\text{Pe}_{\text{s}} = 2016$ (Figures \ref{fig:freezing_stats}c and \ref{fig:freezing_stats}d)
 despite freezing times being sufficiently low, which also show the fact that extra energy is required to overcome the latent heat of freezing.  
\begin{figure}
    \centering
    \begin{picture}(250,150)
    \put(25,0){\includegraphics[width=0.4\textwidth]{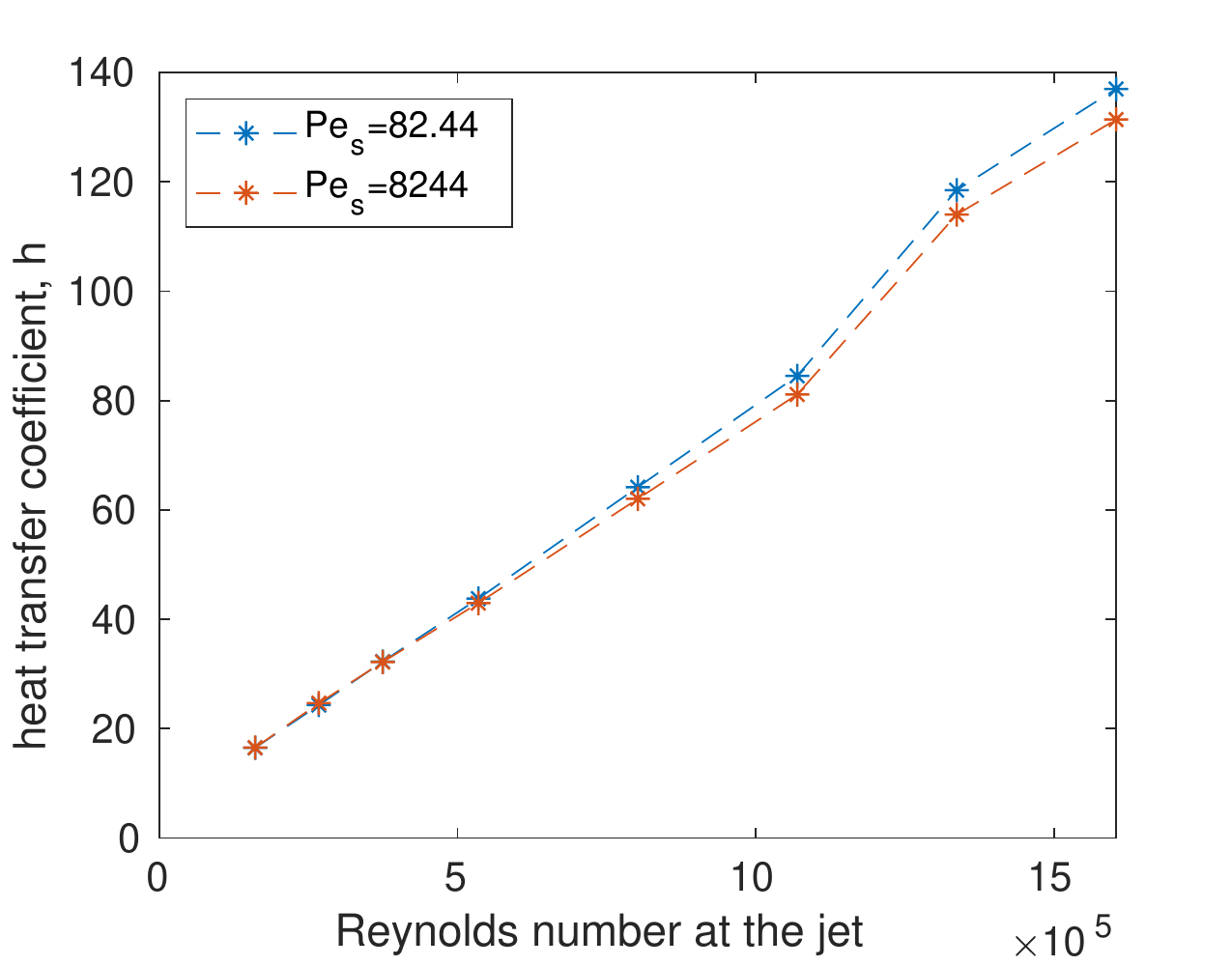}}
    \end{picture}
    \caption{Heat transfer coefficient  $h$ [W/m$^2$K] as function of  Reynolds number for two different values of the  P\'eclet number. The data refer to the domain of H=1.8D.}
    \label{fig:h_coeffs}
\end{figure}
\begin{figure}
    \centering
    \begin{picture}(250,290)
    \put(-15,200){\includegraphics[width=0.55\textwidth]{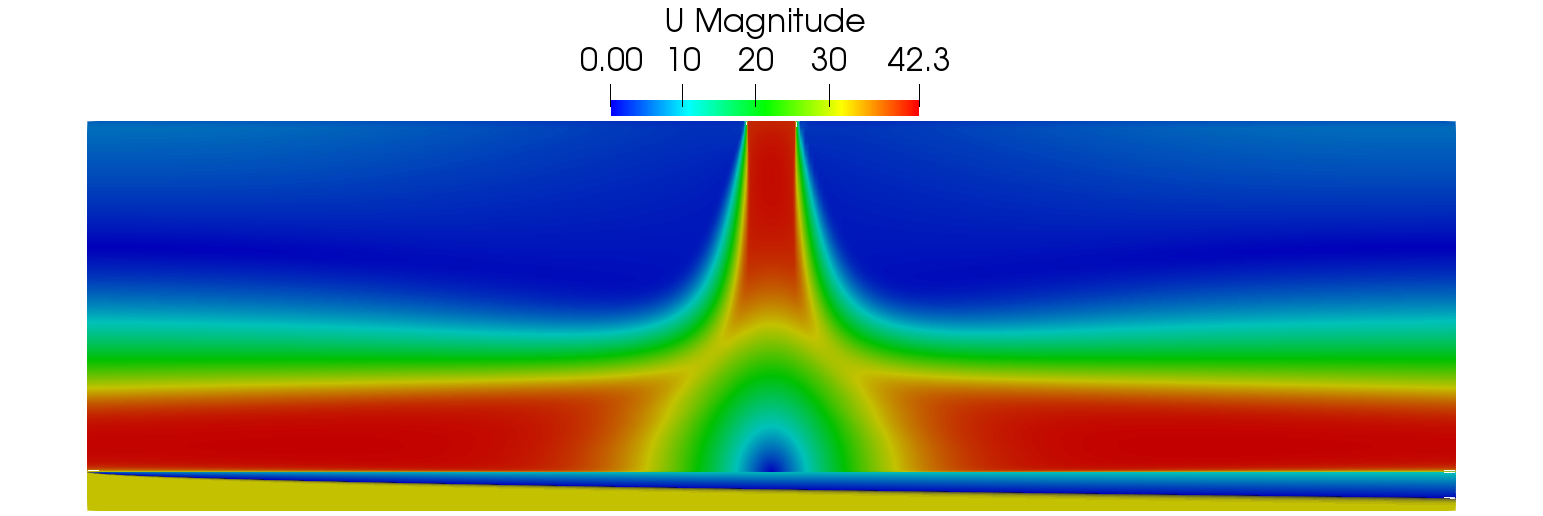}}
    \put(-15,110){\includegraphics[width=0.55\textwidth]{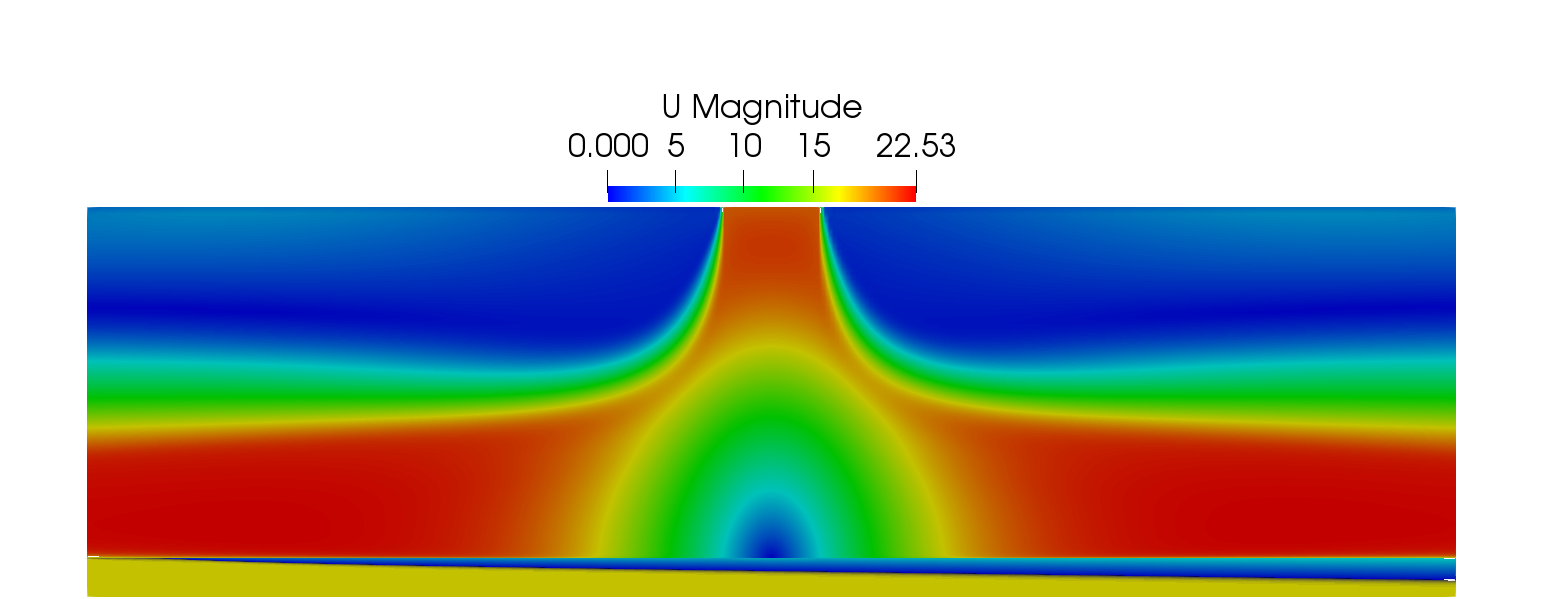}}
    \put(-15,-20){\includegraphics[width=0.55\textwidth]{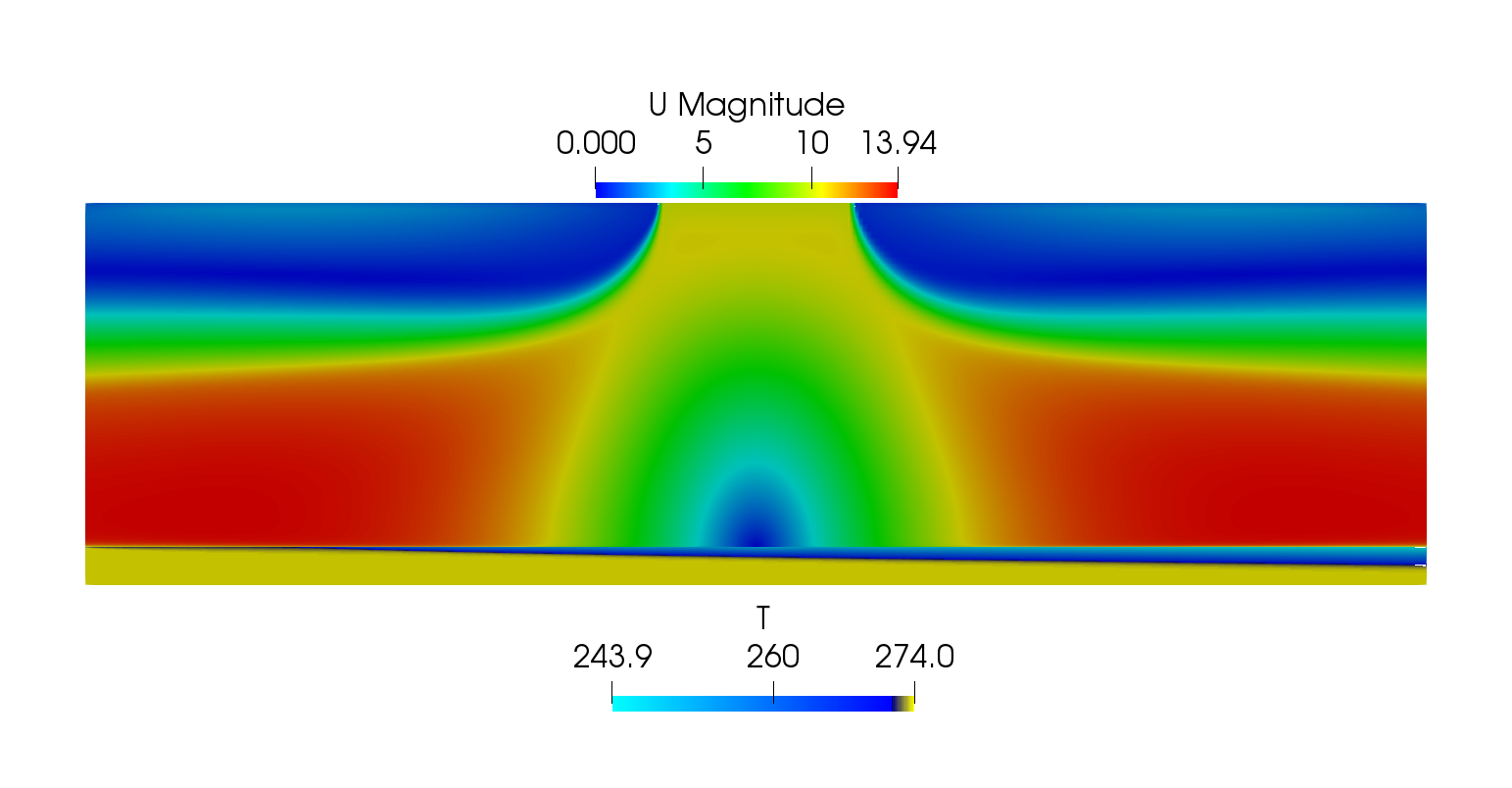}}
    \put(0,275){a)}
    \put(0,185){b)}
    \put(0,95){c)}
    \end{picture}
    \caption{Three computational domains at $\text{Re}=5.348\times10^5$, $\text{Pe}_{\text{s}}=82.44$ using jet diameter to distance to the solid ratios of: a) $\text{H}/0.25\text{D}$ b) $\text{H}/0.5\text{D}$ c) H/D.}
    \label{fig:jet_ratios}
\end{figure}
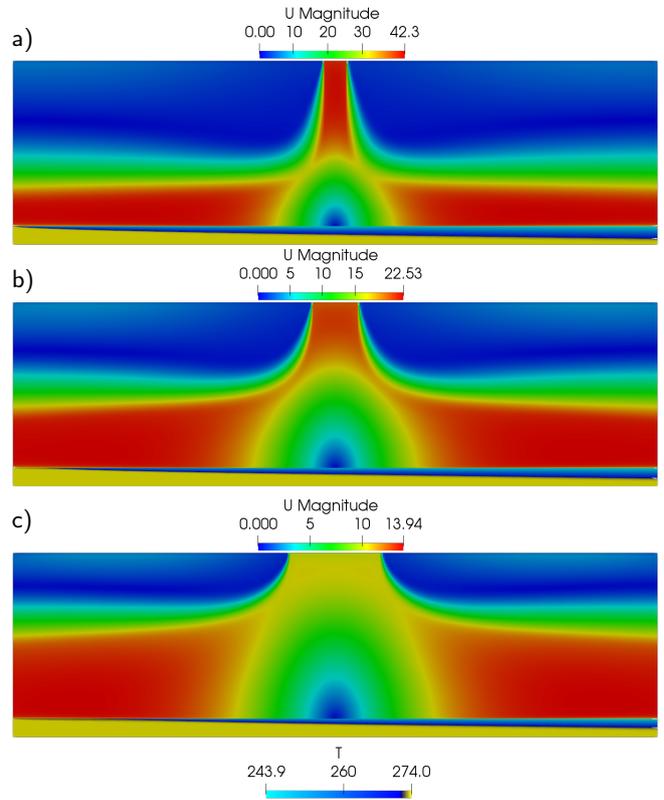

Heat transfer coefficient for the lowest and highest P\'eclet numbers considered ($\text{Pe}_{\text{s}}=82.44$ and $\text{Pe}_{\text{s}}=8244$) is shown in Figure \ref{fig:h_coeffs}. Interestingly, the conveyor speed does not seem to have an appreciable effect on the heat transfer.
This result is quite counter intuitive, since one would think that the additional shear created at the fluid-solid interface should increase fluid mixing and consequently increase the heat transfer coefficient. 
However, it can be explained by the low thermal conductivity of the meat influencing the results.
For higher values of the Reynolds number a small decrease in $h$ can be observed together with the increase of the P\'eclet number. This effect is related to the difference in frozen crust temperature between the two operating conditions.
The effect of quick freezing discussed above can also be seen in Figure \ref{fig:h_coeffs} at the two highest Reynolds numbers, where a a change of slope takes place, when the latent heat of fusion at the solid-fluid interface is overcome.

The just discussed results also show that a more detailed analysis is necessary to investigate the  two dominant parameters of the impingement freezing domain: jet distance from the solid (H) and solid material thickness (H1) which are  explored using the two limiting values of  the P\'clet number ($\text{Pe}_{\text{s}}=82.44$ and $\text{Pe}_{\text{s}}=8244$).

\subsection{Influence of the jet diameter}

The effects of jet distance from the solid have been considered since \cite{sarkar2004modeling}, that reported an optimum jet diameter to distance ratio of 6-8.

In this subsection the computational domain is modified  by reducing the jet diameter two and four times (D$_1$=0.5D and D$_2$=0.25D), effectively obtaining H=3.6D$_1$ and H=7.2D$_2$. 
It should be noted, that all the other geometrical parameters were left identical to the original domain. This allows to keep a solution similarity. 

Simulation results from the three domains are shown in Figure \ref{fig:jet_ratios} and suggest that, for the same value of the Reynolds number, the further away the jet is from the solid domain the quicker is the freezing process. 
Additionally, the fluid bulk region develops between the interface and the jet inlet, which is larger for larger values of D.
This in turn increases the fluid mixing and heat transfer close to the fluid-solid surface, which results in more efficient freezing. 
\begin{figure}
    \centering
    \begin{picture}(250,150)
    \put(25,0){\includegraphics[width=0.4\textwidth]{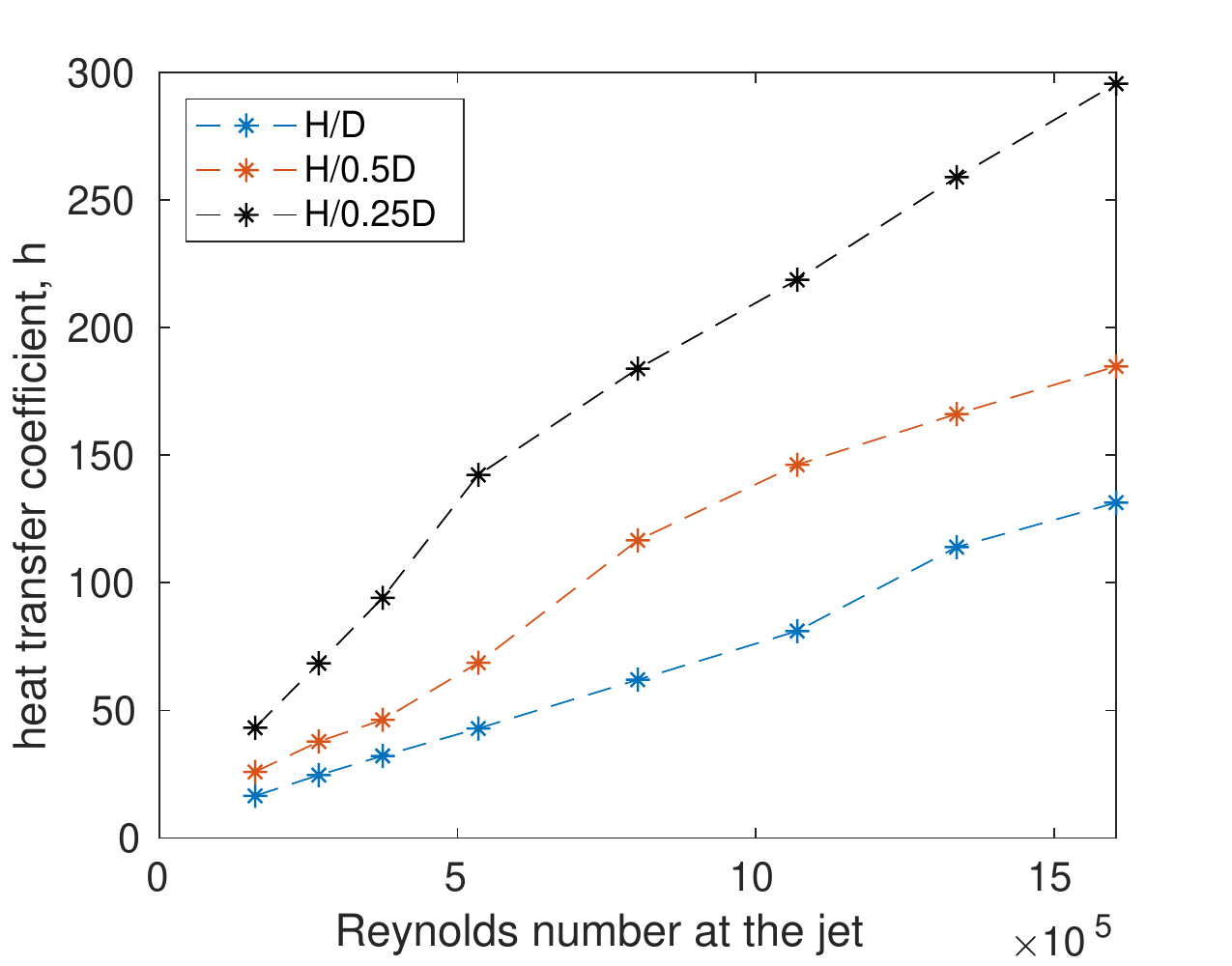}}
    \end{picture}
    \caption{Heat transfer coefficient  $h$ [W/m$^2$K] at  $\text{Pe}_\text{s}=82.44$ using different jet diameter to the distance from the solid.}
    \label{fig:h_coeffs_multiple}
\end{figure}
In terms of heat trasnfer coefficient, this effect is shown in  Figure \ref{fig:h_coeffs_multiple}. The largest heat transfer coefficient results from moving the jet inlet away from the solid in agreement with results from \cite{sarkar2004modeling}. 
In particular, the heat transfer coefficients in the case of H=7.2D is more than doubled than for the H=1.8D domain.
Comparing the freezing time and crust thickness (Figure \ref{fig:freezing_stats-jets}), it can be noticed that increasing the distance of the jet from the solid results in both faster freezing time and deeper crust formation. 
A further increase of the conveyor velocity to the maximum P\'eclet number considered ($\text{Pe}_\text{s}=8244$) results in a similar behaviour to the original domain, and thus in the formation of a very thin frozen crust despite an almost instant freezing time. 
However, for both H=3.6D and H=7.2D the crust is still significantly larger with respect to the case where H=D. This last fact opens the possibility to optimise the freezing process  by just changing the position of the impinging jet.
However, an interesting nonlinear behaviour can be seen at H=7.2D and $\text{Pe}_\text{s}=82.44$ (Figure \ref{fig:freezing_stats-jets}d) where a stabilisation in frozen crust thickness is observed at the highest Reynolds numbers.
This stabilisation indicates that for a specific solid thickness and conveyor velocity, we can derive a critical \textit{Reynolds jet-to-solid} distance over which the impingement effect sharpens and results in an inefficient cooling.

\begin{figure}
    \centering
    \begin{picture}(250,100)
    \put(-15,-20){\includegraphics[width=0.55\textwidth]{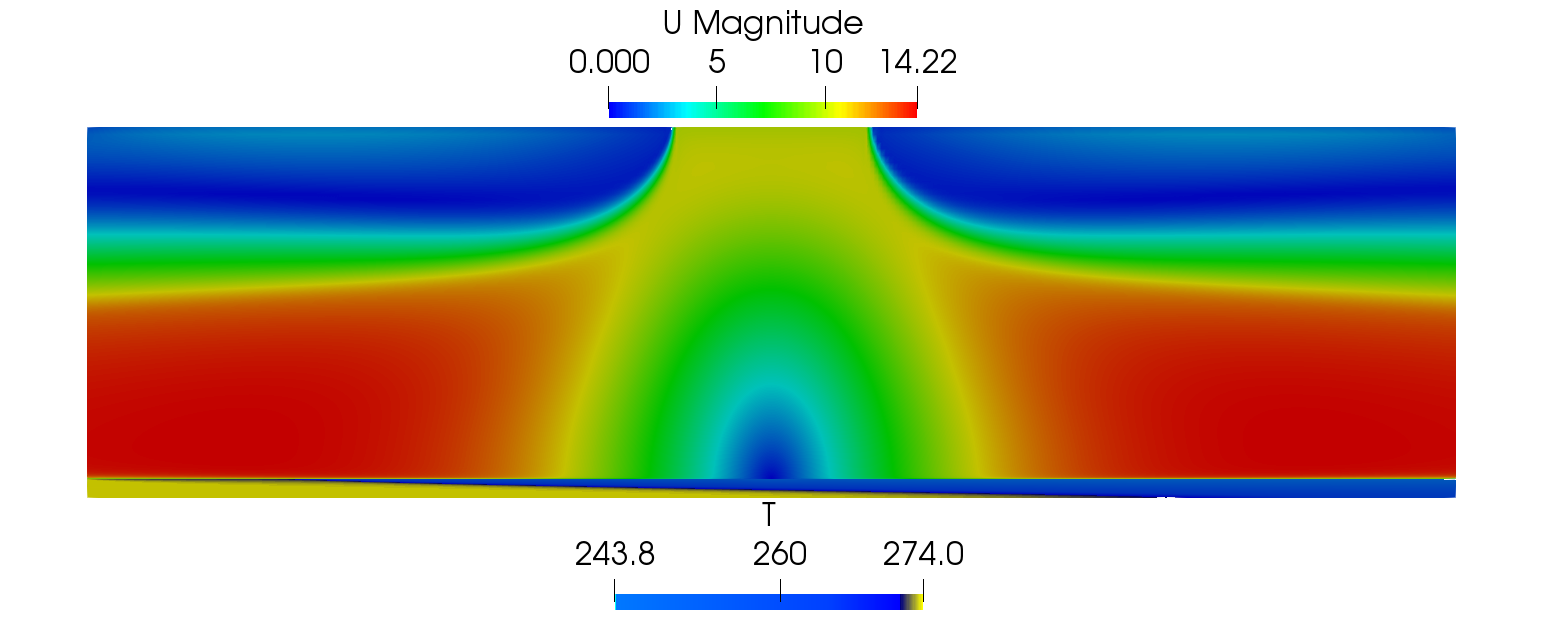}}
    \end{picture}
    \caption{Impingement freezing result at $\text{Re}=5.348\times10^5$, $\text{Pe}_\text{s}=82.44$ using a 50\% thinner solid material domain using a jet diameter to solid distance ratio of H=3.6D.}
    \label{fig:thin_meat}
\end{figure}
\subsection{Influence of the solid diameter}

In this section halving the solid domain diameter (H1) compared to the previous cases is undertaken to study its influence to the freezing  performance. Figure \ref{fig:thin_meat} illustrates the results at $\text{Pe}_\text{s}=82.44$.
\begin{figure}
    \centering
    \begin{picture}(250,170)
    \put(10,0){\includegraphics[width=0.45\textwidth]{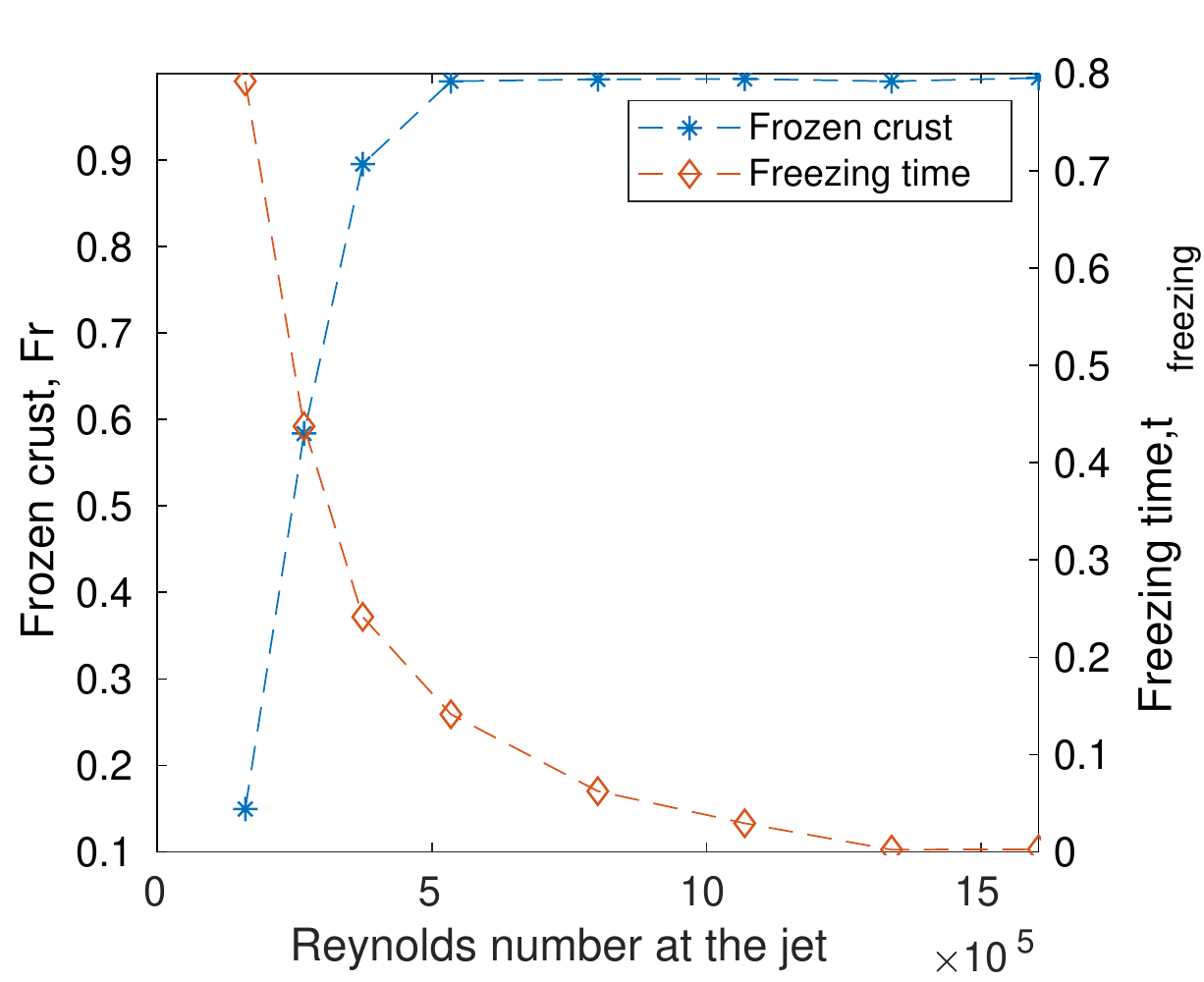}}
    \end{picture}
    \caption{Crust thickness and the freezing time in the half radius domain (H=0.5H$_1$) as a function of the Reynolds number.}
    \label{fig:thin_crust}
\end{figure}
Comparison the results in Figure \ref{fig:jet_ratios}c reveals differences between the domains. 
In the case of a thinner solid, the process is capable of complete freezing (Figure \ref{fig:thin_meat}) with a consistent behaviour across a wide range of Reynolds numbers (Figure \ref{fig:thin_crust}).
Compared to the results in Figure \ref{fig:freezing_stats} at $\text{Pe}_\text{s}=82.44$, freezing is observed to take place at lower Reynolds numbers and at shorter axial coordinates.
This shows a dependence of the freezing characteristics from the solid diameter and the need for process tuning for different products.  

\section{Conclusions}

A numerical model for axial impingement freezing of food products on a moving conveyor is proposed. The model is able to capture phase change, which is critical in capturing the thermophysical description of the freezing process.
The model was implemented in the finite volume open-source library OpenFOAM\textsuperscript{\textregistered} as a custom solver and its numerical convergence is presented in an industrial application.
Key dimensionless parameters were identified to describe the performance of the freezing model. Additionally, a parametric study to investigate the process efficiency under different operating conditions and geometrical configurations of the potential freezing apparatus was undertaken.
The key findings of the study can therefore be summarised by: 
\begin{itemize} 
    \item High Reynolds numbers are required to overcome the effects of the latent heat of freezing. The effect becomes more pronounced at high conveyor speeds, in which the large velocity do not leave sufficient time to the frozen layer to penetrate in depth in the solid.
    \item Freezing time and thickness are dependent on the solid material thickness. By halving the solid domain radius of one half, the complete freezing of the product was found as it moves along the domain. Thus, it requires a significantly lower fluid flow speed and overall heat transfer coefficient. 
\end{itemize}

Future studies may include the investigation of more complicated 3-dimensional geometries as well as the generalisation to heterogeneous and anisotropic food products through the use of improved thermophysical models.  

\section*{Acknowledgements}
The authors would like to thank University of Nottingham Hermes fund for sponsoring the research. 

%
%

\bibliographystyle{unsrt}
\bibliography{cas-refs}
\end{document}